\documentclass[%
 aip,
 amsmath,amssymb,
 reprint,%
]{revtex4-1}

\usepackage{graphicx}
\usepackage{dcolumn}
\usepackage{bm}

\usepackage[utf8]{inputenc}
\usepackage[T1]{fontenc}
\usepackage{mathptmx}
\usepackage{etoolbox}
\usepackage{xcolor}

\makeatletter
\def\@email#1#2{%
 \endgroup
 \patchcmd{\titleblock@produce}
  {\frontmatter@RRAPformat}
  {\frontmatter@RRAPformat{\produce@RRAP{*#1\href{mailto:#2}{#2}}}\frontmatter@RRAPformat}
  {}{}
}%
\makeatother
\begin{document}

\preprint{AIP/123-QED}

\title[Continuum-statistical dynamics of colloidal suspensions under kinematic reversibility]{Continuum-statistical dynamics of colloidal suspensions under kinematic reversibility}
\author{Jerome Burelbach}
\affiliation{International School Michel Lucius, 157 avenue Pasteur, 2311 Luxembourg, Luxembourg}
\email{jerome.burelbach@education.lu}

\date{\today}
	
	\begin{abstract}
    	A continuum-statistical linear-response theory is presented that clarifies the continuum-mechanical origin of Onsager reciprocity in colloidal motion. By decoupling hydrostatic and hydrodynamic stress, it is shown that Onsager reciprocal relations emerge when the Lorentz reciprocal theorem is applied under kinematic reversibility to the auxiliary flow problem of colloidal sedimentation. The theory applies to suspensions containing multiple species of microparticles and derives all non-equilibrium contributions to colloidal diffusion from a single application of the Lorentz reciprocal theorem, irrespective of whether a slip or no-slip hydrodynamic boundary condition is imposed at the colloidal surface. Furthermore, a boundary-layer treatment is assumed only for ciliary motion, while the hydrostatic forces giving rise to non-equilibrium thermodynamic motion are fully resolved beyond the boundary-layer approximation. As an application, a closed-form expression is derived for the diffusiophoretic mobility of a colloid interacting with an ideal solute via a step potential of arbitrary range and strength. The resulting expression recovers established boundary-layer results, including curvature corrections, strong adsorption and slip-enhanced diffusiophoresis, and generalizes them to arbitrary layer thickness, potential strength and hydrodynamic surface slippage, showing that the enhancement by slip saturates once the slip length becomes comparable to the colloid radius.
\end{abstract}

    \maketitle

	\section{Introduction}\label{sec:-8}

	In continuum mechanics, a multicomponent system is represented by continuous media that meet at well-defined boundaries. This is achieved by partitioning the system into a fine mesh of elementary volumes of size $dV$, each of which is homogeneous in the thermodynamic limit. Once the local continuum scale of $dV$ is identified, the dynamics is governed by conservation laws in the form of differential continuity equations for mass, particle number, momentum and energy, expressed in terms of fluxes and source densities defined throughout these media. These continuity equations are closed by assuming that each elementary volume is in local thermodynamic equilibrium (LTE), yielding linear constitutive relations between fluxes and gradients of certain thermodynamic and hydrodynamic continuum fields.\cite{de2013non} Within Onsager’s theory of non-equilibrium thermodynamics, time-reversal symmetry implies that the transport coefficients in these relations are symmetric, a result known as the Onsager reciprocal relations.\cite{onsager1931reciprocal,onsager1931reciprocal2}

    Although kinetic and non-equilibrium statistical theories provide a means to evaluate these transport coefficients from microscopic fluctuations and corresponding correlation functions, such approaches can become computationally expensive for colloidal motion, where a large number of molecules is required to model the local dynamics inside a representative volume element (RVE) of a colloidal suspension. To overcome this bottleneck, Derjaguin instead modeled the suspended microparticles as continuous objects, thus obtaining continuum-mechanical expressions of colloidal transport coefficients based on Onsager reciprocal symmetry.\cite{churaev2013surface} This approach reduces the phenomenological character of coupling in the low Knudsen number regime, where the molecular effect on transport is captured by well-defined boundary conditions. Unlike the Newtonian equations of motion, however, the continuity equations of continuum mechanics are in general not time-reversible. Consequently, the validity of Onsager reciprocity at the continuum scale is simply assumed without proof by such non-equilibrium thermodynamic approaches, which therefore fail to clearly identify the solvent's role in momentum relaxation.\cite{agar1989single,burelbach2018unified} From a continuum-mechanical perspective, such a reciprocal symmetry is justified by the principle of kinematic reversibility, which is a key ingredient in the application of the Lorentz reciprocal theorem.\cite{happel2012low,masoud2019reciprocal} In contrast, continuum-mechanical reciprocal approaches often treat the underlying driving forces phenomenologically within a boundary-layer approximation (BLA), based on an effective slip velocity\cite{anderson1989colloid} or interfacial tension gradient,\cite{masoud2014reciprocal} despite the fact that these may be related to thermodynamic equations of state. Indeed, accurately predicting phoretic mobilities requires fully resolving the coupling between thermodynamic forces and hydrodynamic flows beyond the BLA, especially when the colloidal radius is comparable to the interfacial layer thickness.\cite{burelbach2019linear, de2019self, mayer2023thermophoretic} Hence, reducing the phenomenological character of colloidal transport remains a fruitful area of active research, as evidenced by applications of the Lorentz reciprocal theorem, ranging from Teubner's early treatment of electrophoresis to more recent work, which often focus on force-free transport of a single colloid while assuming a no-slip hydrodynamic boundary condition at its surface.\cite{teubner1982motion,witten2019adapting,marbach2020local,brady2021phoretic,ganguly2024unified}

    The main objective of this work is to derive a general expression for the colloidal diffusion flux in terms of continuum fields, using a reciprocal approach based on the auxiliary flow problem of sedimentation. The resulting framework provides a clear perspective on the underlying assumptions of kinematic reversibility, the solvent’s role in momentum relaxation, and the driving forces governing diffusion and advection in colloidal suspensions. Moreover, it shows that applying the Lorentz reciprocal theorem to an RVE of the suspension, rather than to the fluid region alone, yields all contributions to colloidal diffusion within a single framework, without restriction to a no-slip hydrodynamic boundary condition or to a boundary-layer treatment of the driving forces.

	For a systematic derivation of results, the terminology and assumptions used in colloidal suspensions, including an explicit definition of the condition of representative thermodynamic equilibrium (RTE), are first outlined in Section \ref{sec:-5}, followed by an introduction of disturbance fields and continuity equations in Section \ref{sec:-6}. The core theoretical developments are then presented in Section \ref{sec:-7}, which consists of three subsections. Section \ref{sec:-4} presents the decomposition into hydrodynamic and hydrostatic stress and its relation to thermodynamic forces, and identifies the part of the hydrostatic force density that is balanced by the solvent. In Section \ref{sec:-3}, this decomposition is then used in the application of the Lorentz reciprocal theorem based on the auxiliary flow of colloidal sedimentation, yielding the Onsager reciprocal relations, as given by Eqs. (\ref{eq:-104}) and (\ref{eq:-105}), and a general reciprocal form of the colloidal diffusion velocity, given by Eq. (\ref{eq:-94}). In Section \ref{sec:-1}, the colloidal diffusion velocity is then recast into a computable form in terms of fluid-mechanical disturbances only, and the assumptions entering this form are stated explicitly. The result is given by Eq. (\ref{eq:-100}) and highlights the role of the osmotic force density and the colloidal osmotic pressure gradient in colloidal diffusion.

    The framework is validated in Section \ref{sec:-3} by recovering the swimming velocity due to ciliary motion, which is generalized to arbitrary hydrodynamic surface slippage. In Section \ref{sec:-2}, it is first compared with previous treatments of phoretic motion. In doing so, an earlier prediction of phoretic motion of a single colloid in the absence of an interfacial excess is shown to be spurious. The framework is then applied to diffusiophoresis caused by a step potential between the colloid and an ideal solute. The resulting closed-form expression for the diffusiophoretic velocity recovers established results as limiting cases, generalizes them to arbitrary layer thickness, potential strength and hydrodynamic surface slippage, and shows that the enhancement of diffusiophoresis by slip saturates once the slip length becomes comparable to the colloid radius.

    \section{Preliminaries}
    
	\subsection{Colloidal suspensions: Definitions and assumptions}\label{sec:-5}
	
\begin{figure}
		\centering
		\includegraphics[width=0.3\textwidth]{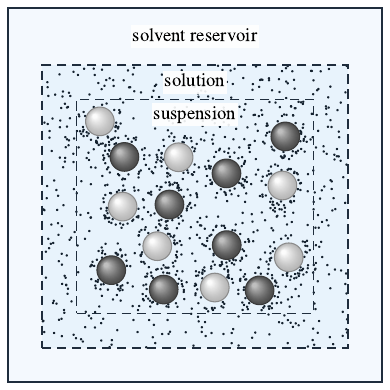}
		\caption{Schematic representation of a suspension. The dark and light gray spheres represent different species of microparticles, which consist of different solid microphases and are suspended in a solution. The solution is a viscous microphase consisting of small solutes (black dots) dispersed in a solvent reservoir (light blue background). If the dark gray spheres are the colloids, then the fluid comprises the solution and the light gray spheres. The dashed lines represent thermodynamic reservoir boundaries.}
		\label{fig:-1}
	\end{figure}

	The system under consideration is a volume-filling, incompressible colloidal suspension, which may include chemical or thermal sources, but must conserve mass internally. Furthermore, the presented framework assumes magnetic fields to be absent, and seeks the translational velocity of the colloids, defined as the volume-averaged velocity of their solid microphase, which coincides with the mean velocity of their centroids. The colloids are free to rotate, and translation-rotation coupling at the particle level is not eliminated, but instead contained in the auxiliary hydrodynamic problem introduced in Section \ref{sec:-3}.\cite{brenner1967coupling} The rotational velocity of the colloids is not determined here, and the explicit results for isolated spherical colloids, such as the swimming velocity and the diffusiophoretic mobility, are restricted to cases where translation and rotation decouple. The local dynamics is described continuum-mechanically at a microscopic scale, relative to an inertial laboratory frame. At this scale, a collection of indistinguishable particles of a given chemical species is referred to as a component of the suspension, and a microphase is defined as a continuous medium composed of a combination of such components. As shown in Fig. \ref{fig:-1}, the suspension consists of different species of microparticles and small solutes dispersed in a solvent reservoir. Within an RVE, represented by a macroscopic region $\mathcal V$ of volume $V=L^3$ chosen large enough to recover the thermodynamic limit of a homogeneous suspension, we have $N_c\ll N_k\ll N_l$, where $N_c$, $N_k$ and $N_l$ denote the number of microparticles of a given species $c$, solute particles of component $k$, and solvent particles of component $l$, respectively.
	
	The solvent and the small solutes dispersed in it constitute a viscous microphase referred to as the solution, denoted by a subscript "$\eta$" and occupying a partial region $\nu_\eta$ within $\mathcal V$, whereas the microparticles are denoted by a subscript "m" and occupy a partial region $\nu_\text{m}=\mathcal V\setminus\nu_\eta$ of volume $V_\text{m}$. Each microparticle species consists of an incompressible solid microphase, made of chemically bound components that cannot diffuse. The microparticles are rigid, such that they can only undergo rigid-body motion and the strain rate tensor $\bm e$ (defined in Section \ref{sec:-4}) vanishes inside them. The solution meets the solid surface of a microparticle at a well-defined interfacial boundary where hydrodynamic boundary conditions apply (see Section \ref{sec:-3}) and where components may be subjected to local body forces deriving from specific interfacial potential interactions. The solution is further assumed to be isotropic, implying that orientational correlations due to interfacial molecular crowding are ignored. The colloids ("$c$") are a chosen species of microparticles whose motion we seek to determine, occupying a partial region $\nu_c$ of volume $V_c$, such that $\nu_\text{m}=\bigcup_c\nu_c$ and $V_\text{m}=\sum_c V_c$, where the union and the sum run over all microparticle species. The fluid, which is denoted by a diacritic "$\sim$", refers to the surrounding solution and all other microparticle species suspended in it, excluding the colloids and thus occupying a partial region $\tilde\nu=\mathcal V\setminus\nu_c$.
	
	The principle of kinematic reversibility, which refers to the reversal of flow fields under reversed driving forces, is based on a low Knudsen number, low Reynolds number, and low Mach number, which are all tied to the solvent's role in momentum relaxation. The solvent is therefore modeled as a strongly coupled, viscous, molecular medium, rapidly responding to non-equilibrium forces to maintain a local mechanical equilibrium under the constraint of incompressible flow. Additionally, kinematic reversibility requires that the continuum-mechanical coupling between flows and forces is independent of the forces. This requirement, which is essential for a linear-response theory based on the Lorentz reciprocal theorem, does not imply spatial uniformity of the corresponding coupling quantities, such as the dynamic viscosity or component number densities, but rather that these quantities remain unchanged under different force fields. Although the microparticles undergo Brownian motion, momentum relaxes on a timescale much shorter than the timescale over which their configuration evolves. On the latter timescale, known as the Smoluchowski timescale, the local flow and stress fields are therefore determined, for any instantaneous configuration of the microparticles, by a deterministic creeping-flow problem to which kinematic reversibility applies.\cite{batchelor1976brownian} The stochastic evolution of this configuration only enters via statistical averages over configurations, such that thermal fluctuations are represented by equilibrium thermodynamic quantities, such as the colloidal osmotic pressure, rather than by fluctuating forces.\cite{batchelor1976brownian}

	The linear-response theory developed here further requires an extension of LTE from the continuum scale of $dV$ to the macroscopic scale of an RVE. This condition, referred to as representative thermodynamic equilibrium (RTE), comprises the following assumptions: (i) the suspension is statistically homogeneous within an RVE, such that ensemble averages can be evaluated as volume averages over $\mathcal V$; (ii) in the absence of non-equilibrium forces, the entire RVE, including all interfacial boundaries, is at thermodynamic equilibrium, such that the equilibrium component distributions follow from equilibrium statistical mechanics; (iii) the fluxes are linear in the non-equilibrium forces, with coupling quantities, such as the dynamic viscosity, the component number densities and the hydrodynamic slip length, evaluated at equilibrium; and (iv) local continuum fields relax on a timescale short compared to the macroscopic timescale over which boundaries change position. The condition of RTE justifies a decomposition into hydrostatic and hydrodynamic stress and implies that the colloidal osmotic pressure derives from a thermodynamic equation of state.\cite{dhont2004thermodiffusion} By virtue of condition (iv), the local continuity equations can moreover be solved at steady state within a given RVE under slowly evolving boundary conditions. In the dilute limit, when all microparticle species are present at low volume fractions, a single colloid suspended in an unbounded solution can then be considered. The term "first order" will be used to refer to quantities evaluated to first order in the forces that arise from non-equilibrium boundary conditions, and equilibrium quantities evaluated to zeroth order in these non-equilibrium forces will be denoted by a superscript "$\text{eq}$".
	
	\subsection{The quantitative tools of continuum statistics: Disturbance fields and continuity equations}\label{sec:-6}
	
	\begin{figure}
		\centering
		\includegraphics[width=0.3\textwidth]{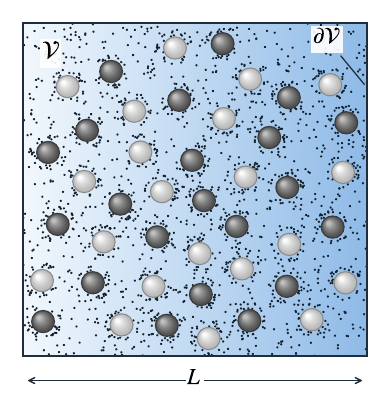}
		\caption{An RVE of a homogeneous suspension. The blue background gradient shows that the volume element is subjected to forces due to non-equilibrium boundary conditions.}
		\label{fig:-3}
	\end{figure}

    \subsubsection{Disturbance fields}
	
    Describing the stochastic motion of every individual colloid in a suspension can be computationally expensive, especially when multiple species of microparticles are present, making the dynamics a challenging many-body interaction problem. However, many industrial applications are not concerned with the motion of each colloid, but rather with the average dynamics of an ensemble of colloids.\cite{batchelor1976brownian} Therefore, we seek a macroscopic, continuum-statistical description of colloidal motion in terms of disturbance fields.
    
    Colloidal diffusion is defined as the disturbance relative to the average motion, or advection, of the suspension. Unlike diffusion in molecular mixtures, which is described relative to the barycentric reference frame,\cite{de2013non} the natural reference frame for colloidal diffusion is the frame of zero mean volume flux, since macroscopic continuity equations can be obtained by averaging over an RVE of a homogeneous suspension.\cite{batchelor1976brownian,brady2011particle} To this end, the system is divided into a mesh of RVEs, each of size $V$. As illustrated in Fig. \ref{fig:-3}, each RVE occupies a macroscopic region $\mathcal V$. By virtue of condition (i) of RTE (Section \ref{sec:-5}), the ensemble average of a local continuum field $\bm q$ over a given RVE at a given time $t$ coincides with its volume average over $\mathcal V$.\cite{brady2011particle} The macroscopic field of $\bm q$, or suspension average, can hence be introduced via
	\begin{equation}
		\left\langle\bm q\right\rangle=\frac{1}{V}\int_{\mathcal V}\bm q\,d^3\bm r.
	\end{equation}
    Throughout this work, spatial derivatives of local fields defined over the entire region $\mathcal V$ are understood in the distributional sense, such that a jump of a field across an internal boundary of $\mathcal V$ gives rise to a surface contribution proportional to a Dirac delta distribution concentrated on that boundary. Gauss' theorem then relates the volume average of $\nabla\cdot\bm q$ to the flux of $\bm q$ across the outer boundary $\partial\mathcal V$, such that, if the RVE size $L$ is small compared to the macroscopic length scale over which $\left\langle\bm q\right\rangle$ varies, we can write
    \begin{equation}
        \left\langle\nabla\cdot\bm q\right\rangle=\bar\nabla\cdot\left\langle\bm q\right\rangle,\label{eq:avg-div}
    \end{equation} 
    where $\bar\nabla$ denotes the macroscopic gradient over adjacent volume elements. If the normal component of $\bm q$ is continuous across boundaries, no surface contributions arise, and $\nabla\cdot\bm q$ coincides with its piecewise divergence evaluated inside each microphase. Similarly, the volume average of the gradient of a scalar field $q$ is given by
    \begin{equation}
        \left\langle\nabla q\right\rangle=\bar\nabla\left\langle q\right\rangle,\label{eq:avg-grad}
    \end{equation}
    where surface contributions vanish if $q$ is continuous across boundaries. Accordingly, the suspension disturbance of $\bm q$ is defined by 
    \begin{equation}
		\bm q'=\bm q-\left\langle\bm q\right\rangle.
	\end{equation} 
	Furthermore, colloidal diffusion is commonly described as driven by fluid-mechanical disturbances relative to the bulk fluid (denoted by a superscript "b"), which fills $\mathcal V$ in the absence of the colloids with the statistical properties of the fluid and is subject to the same non-equilibrium boundary conditions on $\partial\mathcal V$.\cite{batchelor1972sedimentation} The fluid-mechanical disturbance of a local field $\bm q$, induced by the colloids relative to the bulk fluid, is given by 
	\begin{equation}
		\delta\bm q=\bm q-\bm q^\text{b},\label{eq:-103}
	\end{equation}
	where the field $\bm q^\text{b}$ of the bulk fluid is defined everywhere inside $\mathcal V$. The equilibrium field of $\bm q^\text{b}$ will be denoted by $\bm q^{\text{b}|\text{eq}}$. Based on Eq. (\ref{eq:-103}), fluid-mechanical disturbances satisfy $\delta\left(\nabla\cdot\bm q\right)=\nabla\cdot\delta\bm q$, $\delta\left(\nabla g\right)=\nabla\delta g$ and $\delta\left(g\bm q\right)=\delta g\,\bm q+g^\text{b}\,\delta\bm q$, where $g$ is a scalar field.

    \subsubsection{Local and macroscopic continuity equations}
	
	\begin{table}[t]
		\centering
		\renewcommand{\arraystretch}{1.4}
		\begin{tabular}{lcc}
			\hline\hline
			\textbf{Quantity} & \textbf{Local continuum scale} & \textbf{Macroscopic scale} \\
			\hline
			Energy &
			$\displaystyle \nabla\cdot\bm j_q=Q$
			&
			$\displaystyle \bar\nabla\cdot\left\langle\bm j_q\right\rangle=\left\langle Q\right\rangle$
			\\
			
			Momentum &
			$\displaystyle \bm f+\nabla\cdot\bm\sigma=0$
			&
			$\displaystyle \left\langle\bm f\right\rangle+\bar\nabla\cdot\left\langle\bm\sigma\right\rangle=0$
			\\
			
			Particle number &
			$\displaystyle \nabla\cdot\bm J_i=\Gamma_i$
			&
			$\displaystyle \partial_t\left\langle n_i\right\rangle+\bar\nabla\cdot\left\langle\bm J_i\right\rangle=\left\langle\Gamma_i\right\rangle$
			\\
			\hline\hline
		\end{tabular}
		\caption{Local and macroscopic continuity equations for energy, momentum and particle number in a colloidal suspension. The fluxes $\bm j_q$, $\bm\sigma$ and $\bm J_i$ correspond to the local heat diffusion flux, stress tensor and particle flux of component $i$, respectively. $\bm f$ is the local body force density, whereas the local source densities $Q$ and $\Gamma_i$ account for the production or consumption of heat and particles of component $i$, respectively. The term $\partial_t\left\langle n_i\right\rangle$ is the time derivative of the volume average of $n_i$, describing the macroscopic evolution of component $i$. By virtue of condition (iv) of RTE, the local continuity equations are given in their steady-state forms.}
		\label{tab:continuity_equations}
	\end{table}
	
	Based on the assumed properties of colloidal suspensions introduced in Section \ref{sec:-5}, the forms of the local and macroscopic continuity equations for energy, momentum and particle number are summarized in Table \ref{tab:continuity_equations}. Note that the local continuity equations hold everywhere inside $\mathcal V$, whether inside the solution or inside the microparticles. Since their derivatives are understood in the distributional sense, a jump in the normal component of a local flux across a boundary must be balanced by a source density concentrated on that boundary. Such interfacial sources, for instance due to surface reactions or to body forces deriving from discontinuous interfacial potentials, are included in $Q$, $\Gamma_i$ and $\bm f$ as surface densities. In their absence, the normal components of $\bm j_q$ and $\bm J_i$, as well as the traction $\bm\sigma\cdot\bm n$, are continuous across boundaries, where $\bm n$ is the unit normal vector. Inside the microparticles, the local flow velocity $\bm u$ introduced below corresponds to the rigid-body velocity field of each microparticle. 
    
    In colloidal suspensions, energy transport occurs primarily via heat conduction, which is much faster than particle transport. At both the local and macroscopic scales, conservation of energy can therefore be described using the steady-state form of the heat equation, governed by the heat diffusion flux $\bm j_q=-\kappa\nabla T$, which couples to the gradient in temperature $T$ via the thermal conductivity $\kappa$. Here, the time derivative is negligible by virtue of condition (iv) of RTE, and convective heat transport by the local flow is negligible at low thermal Péclet number. On the Smoluchowski timescale introduced in Section \ref{sec:-5}, inertia is negligible and the suspension is at mechanical equilibrium, such that the local and macroscopic momentum continuity equations are respectively given by 
    \begin{eqnarray}
		&&\bm f+\nabla\cdot\bm\sigma=0,\label{eq:-151}\\
		&&\left\langle\bm f\right\rangle+\bar\nabla\cdot\left\langle\bm\sigma\right\rangle=0,
	\end{eqnarray} 
    and solved under the constraints of incompressible flow, expressed by
	\begin{eqnarray}
		&&\nabla\cdot\bm u=0,\label{eq:-23}\\
		&&\bar\nabla\cdot\left\langle\bm u\right\rangle=0,
	\end{eqnarray}
	where $\bm u$ is the local volume-centered flow velocity and $\left\langle\bm u\right\rangle$ is the advective suspension velocity. The particle flux $\bm J_i$ can be expressed as $\bm J_i=n_i\bm u+\bm j_i$, where $n_i$ is the number density of component $i$. The local diffusion flux $\bm j_i$ describes motion of the particles of component $i$ relative to the flow velocity $\bm u$. If the index $i$ refers to a component of a solid microphase (denoted by $i\in\nu_\text{m}$), where diffusion is prohibited, we simply have $\bm j_i=0$. The local diffusion fluxes satisfy $\sum_i \bar v_i\bm j_i=0$, where $\bar v_i$ is the partial volume per particle of component $i$.	
	
	To describe colloidal diffusion, namely motion relative to advection, we must introduce the decomposition 
    \begin{equation}
        \bm u=\left\langle\bm u\right\rangle+\bm u',
    \end{equation}
    where $\bm u'$ is the diffusion flow velocity, such that $\left\langle\bm u'\right\rangle=0$ and $\nabla\cdot\bm u'=0$. Hence, the local and macroscopic particle fluxes of component $i$ can be written as $\bm J_i=n_i\left\langle\bm u\right\rangle+n_i\bm u'+\bm j_i$ and $\left\langle\bm J_i\right\rangle=\left\langle n_i\right\rangle\left\langle\bm u\right\rangle+\bm{\mathcal J}_i$, where $\bm{\mathcal J}_i=\left\langle n_i\bm u'\right\rangle+\left\langle\bm j_i\right\rangle$ is the macroscopic diffusion flux and $\left\langle n_i\bm u'\right\rangle$ is the macroscopic diffusion flow of component $i$. By virtue of condition (iv) of RTE, the time derivative of the local number density $n_i$ can be neglected. Moreover, since local diffusion is fast compared to diffusion flow inside the solution ($|\bm j_i|/n_i\gg|\bm u'|$ for $i\in\nu_\eta$, corresponding to a low local Péclet number), convection by the diffusion flow velocity $\bm u'$ is negligible, and the local conservation of particle number for the solute and solvent components reduces to a steady-state equation given by 
    \begin{equation}
        \nabla\cdot\bm j_i=\Gamma_i
    \end{equation} 
    inside the advective frame of zero mean volume flux. As a result, the thermodynamic forces that drive these diffusion fluxes have no explicit dependence on the local flow velocity field, as discussed by Teubner in the context of electrophoresis.\cite{teubner1982motion}
    
    To introduce the colloids as a macroscopic component of the suspension, we consider $N_c$ identical colloids of species $c$ that occupy a partial region $\nu_c$ of the region $\mathcal V$. Note that the fluid surrounding these colloids may comprise other species of microparticles. The number of constituent component particles per colloid can be written as $\mathcal N_c=V\sum_j\left\langle n_j\right\rangle/N_c$, where the index $j$ runs over all components that are part of the solid microphase of the colloids. Assuming that the constituent components of the colloids are conserved ($\Gamma_j=0$), such that chemical activity at the colloidal surface only involves components of the fluid, $\mathcal N_c$ is constant and the number density of the colloidal component can be defined as $n_c=\sum_j\left\langle n_j\right\rangle/\mathcal N_c$, and the colloidal particle flux as $\bm J_c=\sum_j \left\langle\bm J_j\right\rangle/\mathcal N_c=n_c\left\langle\bm u\right\rangle+\bm{\mathcal J}_c$. Summing the macroscopic continuity equations of the constituent components and dividing by $\mathcal N_c$, the colloidal continuity equation becomes
	\begin{equation}
		\frac{\partial n_c}{\partial t}+\left\langle\bm u\right\rangle\cdot\bar\nabla n_c+\bar\nabla\cdot\bm{\mathcal J}_c=0.\label{eq:-150}
	\end{equation}
	A creation or annihilation of colloids as whole objects, for instance by nucleation or aggregation, would simply require an additional source term in Eq. (\ref{eq:-150}), which is not considered here. Since components of the colloidal microphase cannot undergo local diffusion, and since the number density $\sum_j n_j$ of constituent components is uniform inside the homogeneous, incompressible solid microphase of the colloids, the colloidal diffusion flux is a pure diffusion flow, given by 
    \begin{equation}
		\bm{\mathcal J}_c=n_c\left(\left\langle\bm u\right\rangle_{\nu_c}-\left\langle\bm u\right\rangle\right),\label{eq:-116}
	\end{equation}
    where $\langle\cdot\rangle_{\nu_c}$ denotes the volume average over the region $\nu_c$ of the colloidal microphase. As noted in Section \ref{sec:-5}, $\left\langle\bm u\right\rangle_{\nu_c}$ coincides with the mean velocity of the colloidal centroids.
    
    As outlined in the introduction, our aim is to derive a continuum-statistical expression for the colloidal diffusion flux $\bm{\mathcal J}_c$ using the Lorentz reciprocal theorem and the auxiliary flow of colloidal sedimentation. Section \ref{sec:-4} will therefore focus on the decoupling of forces that drive flow from those forces that respond to flow under the constraint of incompressibility, which is a prerequisite for a successful application of this reciprocal approach.
	
	\section{Core theoretical advances: A continuum-statistical description of colloidal diffusion}\label{sec:-7}
    
    \subsection{Hydrostatic forces and hydrodynamic stresses in colloidal suspensions}\label{sec:-4}
	
	To bridge the gap between non-equilibrium thermodynamic and continuum-mechanical reciprocal relations, we must decouple hydrostatic forces from hydrodynamic forces in a suspension. Based on the assumption of RTE, the stress tensor $\bm\sigma$ is therefore decomposed into a hydrostatic part $\bm\sigma_\text{s}$, and a hydrodynamic part $\bm\sigma_\text{d}$ that vanishes in the absence of flow (when $\bm u=0$ everywhere): 
	\begin{equation}
        \bm\sigma=\bm\sigma_\text{s}+\bm\sigma_\text{d},\label{eq:-101}
	\end{equation} 
	where the letters "s" and "d" stand for "static" and "dynamic", respectively. Note that the above decomposition, and hence the assumption of RTE, is often made implicitly. A well-known example is buoyancy: the buoyant force on a microparticle is determined from the hydrostatic pressure field in a static fluid and is assumed to be independent of the velocity gradients that arise during sedimentation. 
    
    The stress tensor can be decomposed into an isotropic part and a deviatoric part:
    \begin{equation}
        \bm\sigma=-P\bm I+\bm\Sigma,
	\end{equation}
    where $P=-\mathrm{Tr}\,\bm\sigma/3$ is the pressure, $\bm I$ is the identity tensor, and $\bm\Sigma$ is the deviatoric stress tensor. The deviatoric stress tensor
    \begin{equation}
        \bm\Sigma=\bm\Sigma_\eta+\bm\Sigma_\text{m}
	\end{equation}
    consists of two separate stress fields, namely a purely viscous shear stress $\bm\Sigma_\eta$ inside the solution and a rigid constraint stress $\bm\Sigma_\text{m}$ inside the microparticles. The constraint stress $\bm\Sigma_\text{m}$ is symmetric, since the microparticles support no couple stresses, and acts as a Lagrange multiplier that enforces the rigidity condition $\bm e=0$ inside the microparticles. Its spatial dependence is therefore not determined by a constitutive relation, which, however, is not needed in what follows.\cite{brady2011particle} The linear constitutive relation for the viscous stress tensor under incompressible flow is given by 
    \begin{equation}
        \bm\Sigma_\eta=2\eta\bm e,
	\end{equation}
    where $\eta$ is the dynamic viscosity of the solution, and $\bm e=(\nabla\bm u+\left(\nabla\bm u\right)^\dagger)/2$.
	Since the microparticles are only in direct contact with the solution and hence under isotropic hydrostatic stress, all deviatoric stresses are of hydrodynamic origin, and the decomposition given by Eq. (\ref{eq:-101}) takes the form
    \begin{eqnarray}
        &&\bm\sigma_\text{s}=-P_\text{s}\bm I,\\
        &&\bm\sigma_\text{d}=-P_\text{d}\bm I+\bm\Sigma_\text{m}+\bm\Sigma_\eta,
    \end{eqnarray} 
    such that $P=P_\text{s}+P_\text{d}$, where $P_\text{s}$ and $P_\text{d}$ are the hydrostatic and hydrodynamic pressure, respectively. 
    Furthermore, the body force density is given by 
    \begin{equation}
        \bm f=\sum_i n_i\left(\bm F_i-\nabla\phi_i\right),\label{eq:-126}
    \end{equation}
    where $\phi_i$ is the interfacial interaction potential of component $i$, resulting only from those effective potential interactions that occur at an interfacial boundary. The non-equilibrium body force $\bm F_i$ acting on component $i$ is a conservative force consisting of an electric force and a gravitational force, such that $\bm F_i=ez_i\bm E+m_i\bm g$, where $e$ is the elementary charge, $z_i$ and $m_i$ are the charge number and mass of a particle of component $i$, and $\bm E$ and $\bm g$ are the corresponding electric and gravitational field vectors. Since $\nabla\phi_i$ is of zeroth order in the non-equilibrium forces, the number density $n_i$ in the interfacial term of Eq. (\ref{eq:-126}) includes first-order deviations from its equilibrium value $n_i^\text{eq}$, whereas $n_i$ can be replaced by $n_i^\text{eq}$ in the first-order term $n_i\bm F_i$. Overall electroneutrality of $\mathcal V$ implies that $\left\langle\omega\right\rangle=0$, where $\omega=\sum_i e z_i n_i^\text{eq}$ is the electric charge density. Since, moreover, interfacial potential interactions cause equal and opposite forces on the microphases inside $\mathcal V$, and the net electric force on an electroneutral RVE vanishes, we have $\left\langle\bm f\right\rangle=\left\langle\rho\right\rangle\bm g$, where $\rho=\sum_i m_i n_i^\text{eq}$ is the mass density. The gravitational field vector $\bm g$ is uniform throughout the suspension, whereas the non-equilibrium electric field $\bm E$ satisfies $\epsilon^\text{eq}\nabla\cdot\bm E=0$ to first order, where $\epsilon$ is the electric permittivity. 

    At RTE, the interfacial boundaries within an RVE are at chemical and thermal equilibrium. Consequently, the interfacial interaction potential $\phi_i$ can be absorbed into the chemical potential $\mu_i$ of component $i$, such that
    \begin{equation}
        \mathring\mu_i=\mu_i+\phi_i.\label{eq:-71}
    \end{equation} 
    The modified chemical potential $\mathring\mu_i$ corresponds to the chemical potential of component $i$ inside its homogeneous microphase when interfacial potential interactions are absent, and is therefore continuous everywhere within the microphase. As shown in Appendix \ref{app:-1}, the Gibbs-Duhem equation can be used to write the hydrostatic pressure gradient to first order as
    \begin{equation}
        \nabla P_\text{s}= \check h^\text{eq}\frac{\nabla T}{T^\text{eq}}+\sum_i n_i^\text{eq}\nabla_{T,P_\text{d}}\mathring\mu_i-\sum_i n_i\nabla\phi_i,\label{eq:-65}
    \end{equation}
    where the subscripts "$T$" and "$P_\text{d}$" of $\nabla$ indicate that the gradient is evaluated at constant temperature and constant hydrodynamic pressure. The modified enthalpy density $\check h$ is defined by $\check h=h-\sum_i n_i \bar H_i$, where $h$ is the net enthalpy density and where $\bar H_i=-T^2\left(\partial\left(\mathring\mu_i/T\right)/\partial T\right)_{\left\{n_j\right\}}$ is the partial enthalpy per particle evaluated at constant component number densities. With Eqs. (\ref{eq:-126})-(\ref{eq:-71}), the hydrostatic force density, defined by
    \begin{equation}
        \bm{\mathcal F}=\bm f-\nabla P_\text{s},\label{eq:-107}
    \end{equation}
    becomes a linear combination of thermodynamic forces with equilibrium scalar densities. Since the interfacial potential gradient terms in Eqs. (\ref{eq:-126}) and (\ref{eq:-65}) are evaluated with the same number densities, they cancel exactly, and all remaining terms are of first order, such that\cite{burelbach2018unified}
	\begin{equation}
		\bm{\mathcal F}=-\check h^\text{eq}\frac{\nabla T}{T^\text{eq}}+\sum_i n_i^\text{eq}\left(\bm F_i-\nabla_{T,P_\text{d}}\mathring\mu_i\right).\label{eq:-88}
	\end{equation}
	The interfacial potential gradients $\left\{\nabla\phi_j\right\}$ therefore do not act as thermodynamic forces, which is a well-known fact for electric double layer forces,\cite{wurger2010thermal} and a direct consequence of the condition of RTE. In particular, $\bm{\mathcal F}$ contains no contributions concentrated on boundaries, since $\mathring\mu_i$ is continuous within each microphase, whereas $\nabla T$ and $\bm F_i$ remain finite.  

    Eq. (\ref{eq:-88}) holds everywhere inside $\mathcal V$. Inside the solution, however, it can be simplified further by considering the role of the solvent in maintaining mechanical equilibrium. Since the local diffusion fluxes are constrained by $\sum_i \bar v_i\bm j_i=0$, the thermodynamic forces in the solution are not all independent, and the modified chemical potential of the solvent is not governed by a transport equation of its own. Instead, the solvent responds to maintain a mechanical equilibrium under the constraint of incompressible flow. To identify the part of $\bm{\mathcal F}$ that can be balanced hydrostatically, namely in the absence of flow, the scalar equilibrium densities inside the solution are decomposed into their values in the homogeneous state of the solution, where $\left\{\phi_j\right\}=0$, and the corresponding equilibrium-interfacial excess densities
    \begin{eqnarray}
        &&\check h^{\phi|\text{eq}}=\check h^{\text{eq}}-\check h^{\text{eq}}\left(\left\{\phi_j\right\}=0\right),\\
        &&n_i^{\phi|\text{eq}}=n_i^\text{eq}-n_i^\text{eq}\left(\left\{\phi_j\right\}=0\right).\label{eq:-156}
    \end{eqnarray}
    In the homogeneous state, the scalar equilibrium densities $\check h^\text{eq}\left(\left\{\phi_j\right\}=0\right)$ and $n_i^\text{eq}\left(\left\{\phi_j\right\}=0\right)$ are uniform. Since $\bm F_i$ is conservative, the corresponding part $\bm{\mathcal F}\left(\left\{\phi_j\right\}=0\right)$ of the hydrostatic force density is then the gradient of a scalar field inside the solution, and hence conservative ($\nabla\times\bm{\mathcal F}\left(\left\{\phi_j\right\}=0\right)=0$). Such a conservative force density cannot drive viscous flow under the constraint of incompressibility,\cite{parola2004particle} but is balanced hydrostatically by the solvent components,\cite{brady2011particle} which adjust their modified chemical potentials such that $\bm{\mathcal F}\left(\left\{\phi_j\right\}=0\right)=0$ everywhere inside the solution. A familiar example is a homogeneous solution subjected to a uniform solute concentration gradient, which remains at rest. Consequently, only the equilibrium-interfacial excess densities contribute to the hydrostatic force density inside the solution, and Eq. (\ref{eq:-88}) reduces to
    \begin{equation}
        \bm{\mathcal F}=-\check h^{\phi|\text{eq}}\frac{\nabla T}{T^\text{eq}}+\sum_{i\in\nu_\eta} n_i^{\phi|\text{eq}}\left(\bm F_i-\nabla_{T,P_\text{d}}\mathring\mu_i\right).\label{eq:-67}
    \end{equation}
    Note that, for a single colloid, the excess densities $\check h^{\phi|\text{eq}}$ and $n_i^{\phi|\text{eq}}$ coincide with the equilibrium fluid-mechanical disturbances $\delta\check h^\text{eq}$ and $\delta n_i^\text{eq}$, whereas they generally differ when other species of microparticles are present, since they then also account for the interfacial layers of these microparticles. Furthermore, the Curie symmetry principle implies that the local diffusion fluxes of an isotropic solution only couple to vectorial thermodynamic forces,\cite{de2013non} such that
	\begin{equation}
		\bm j_i=-L_{iq}\frac{\nabla T}{T^\text{eq}}+\sum_{j\in\nu_\eta}L_{ij}\left(\bm F_j-\nabla_T\mathring\mu_j\right),\label{eq:-130}
	\end{equation}
	where $L_{ij}$ and $L_{iq}$ are the Onsager transport coefficients, which depend on the equilibrium number densities, but are independent of the thermodynamic forces. The diagonal Onsager coefficient can be written as $L_{ii}=n_i^\text{eq}/\xi_i$, where $\xi_i$ is the friction coefficient of a particle of component $i$, whereas the cross-coupling coefficients in Eq. (\ref{eq:-130}) satisfy the Onsager reciprocal relations $L_{iq}=L_{qi}$ and $L_{ij}=L_{ji}$, implying that they also describe the heat flux and diffusion flux of component $j$ caused by a body force on component $i$, respectively. In accordance with Section \ref{sec:-6}, the former contribution to the heat flux, known as the Dufour effect, is neglected, as is common in liquids.
    
	Based on Eq. (\ref{eq:-107}), the local and macroscopic momentum continuity equations now take the forms
	\begin{eqnarray}
		&&\bm{\mathcal F}+\nabla\cdot\bm\sigma_\text{d}=0,\label{eq:-6}\\
		&&\left\langle\bm{\mathcal F}\right\rangle+\bar\nabla\cdot\left\langle\bm\sigma_\text{d}\right\rangle=0.\label{eq:-14}
	\end{eqnarray} 
	Since $\bm{\mathcal F}$ contains no contributions concentrated on boundaries, Eq. (\ref{eq:-6}) implies that the hydrodynamic traction $\bm\sigma_\text{d}\cdot\bm n$ is continuous across all boundaries, whereas a jump in the hydrostatic traction $\bm\sigma_\text{s}\cdot\bm n$ is balanced by a body force density concentrated on the boundary, such as one deriving from a discontinuous interfacial potential. Although the solution of Eq. (\ref{eq:-14}) is not required for a description of colloidal diffusion pursued here, the advective velocity $\left\langle\bm u\right\rangle$ can in principle be obtained from Eq. (\ref{eq:-14}) by relating $\left\langle\bm\sigma_\text{d}\right\rangle$ to the corresponding strain rate tensor $\left\langle\bm e\right\rangle$ via the effective dynamic viscosity of the suspension. Rather, the dynamics of colloidal diffusion is determined by the local momentum continuity equation, whose form in terms of hydrostatic forces and hydrodynamic stresses given by Eq. (\ref{eq:-6}) substantially facilitates the application of the Lorentz reciprocal theorem.  
	
	\subsection{Reciprocal formulation of colloidal diffusion}\label{sec:-3}
	
	In view of Eq. (\ref{eq:-116}), a description of colloidal diffusion thus requires a constitutive relation for the volume-averaged velocity $\bm v=\left\langle\bm u\right\rangle_{\nu_c}$ of an ensemble of $N_c=n_cV$ colloids relative to the advective velocity $\left\langle\bm u\right\rangle$. For a statistically homogeneous suspension, it will now be shown that this is achievable via application of the Lorentz reciprocal theorem by exploiting the auxiliary flow problem of colloidal sedimentation. 
	
	In the auxiliary flow problem of colloidal sedimentation (denoted by a superscript "$*$"), a uniform body force $\bm F^*$ is applied to every colloid, corresponding to a momentum continuity equation $\bm f^*+\nabla\cdot\bm\sigma_\text{d}^*=0$, where $\bm f^*$ takes the value $N_c\bm F^*/V_c$ inside the colloids and zero elsewhere. The flow induced by sedimentation is a diffusion flow, such that
	$\left\langle\bm u^*\right\rangle=0$. Since each colloid sediments at a given velocity that may vary across the ensemble, the mean sedimentation velocity $\bm v^*$ is defined by
	\begin{equation}
		\bm v^*=\left\langle\bm u^*\right\rangle_{\nu_c}.\label{eq:-81}
	\end{equation}
	For any instantaneous configuration of the microparticles, the auxiliary problem is a deterministic creeping-flow problem (Section \ref{sec:-5}), whose flow velocity and hydrodynamic stress are linear in the applied force $\bm F^*$ under kinematic reversibility.\cite{happel2012low} Since the mean sedimentation velocity is itself linear in $\bm F^*$, these fields can be expressed by tensorial relations that are linear in $\bm v^*$, and this linearity is preserved by the subsequent averaging over configurations of a homogeneous suspension.\cite{batchelor1976brownian} We can therefore write
	\begin{eqnarray}
		&&\bm u^*=\bm U^*\cdot\bm v^*,\label{eq:-11}\\
		&&\bm\sigma_\text{d}^*=\bm{\mathcal P}^*\cdot\bm v^*,\label{eq:-12}
	\end{eqnarray}
	where the third-order tensor $\bm{\mathcal P}^*$ is symmetric in its first two indices, since $\bm\sigma_\text{d}^*$ is symmetric. We will refer to $\bm U^*$ and $\bm{\mathcal P}^*$ as the Stokes flow tensor and Stokes stress tensor, respectively. The Stokes flow tensor $\bm U^*$ satisfies 
	\begin{eqnarray}
		&&\nabla\cdot\bm U^*=0,\label{eq:-132}\\
		&&\left\langle\bm U^*\right\rangle=0,\label{eq:-16}\\ &&\left\langle\bm U^*\right\rangle_{\nu_c}=\bm I.\label{eq:-18}
	\end{eqnarray}
	In addition, we have $\bm F^*=\bm\xi\cdot\bm v^*$, where the Stokes friction tensor $\bm\xi$ of the colloids is symmetric and related to the hydrodynamic traction tensor $\bm T^*=\bm n\cdot\bm{\mathcal P}^*$ via $N_c\bm\xi=-\oint_{\partial\nu_c}\bm T^*\,dS$, where $\bm n$ is the unit normal vector pointing out of the corresponding boundaries. Since $\bm f^*$ is uniform inside each colloid, it exerts no torque about the colloidal centroids. Nevertheless, the colloids may rotate in the auxiliary problem, for instance due to their shape or due to hydrodynamic interactions, in which case the corresponding rigid-body rotation is contained in $\bm U^*$. Eq. (\ref{eq:-18}) then remains valid, since rigid-body rotation about the centroid averages to zero over the volume of each colloid, and $\bm\xi$ corresponds to the inverse of the translational mobility of the colloids at zero torque, which is symmetric by virtue of the Lorentz reciprocal theorem.\cite{brenner1967coupling}
	
	To derive the Lorentz reciprocal theorem for an RVE of a colloidal suspension, we first follow the standard procedure of contracting the forces from the local momentum continuity equation of each problem with the flow velocity of the other problem.\cite{happel2012low,lorentz1895attempt} However, rather than using the steady-state Cauchy form of the momentum continuity equation, as given by Eq. (\ref{eq:-151}), we make use of its form in terms of the hydrostatic force density and hydrodynamic stress, given by Eq. (\ref{eq:-6}). Since both contractions are zero, equating them gives
	\begin{equation}
		\bm u\cdot\left(\bm f^*+\nabla\cdot\bm\sigma_\text{d}^*\right)=\bm u^*\cdot\left(\bm{\mathcal F}+\nabla\cdot\bm\sigma_\text{d}\right).\label{eq:-86}
	\end{equation}
	Collecting the terms related to the hydrodynamic stresses $\bm\sigma_\text{d}$ and $\bm\sigma_\text{d}^*$ on the right-hand side, we obtain
	\begin{equation}
		\bm u^*\cdot\bm{\mathcal F}-\bm u\cdot\bm f^*=\bm u\cdot\left(\nabla\cdot\bm\sigma_\text{d}^*\right)-\bm u^*\cdot\left(\nabla\cdot\bm\sigma_\text{d}\right).\label{eq:-68}
	\end{equation}
	Each term on the right-hand side can further be rewritten using the identity $\bm u\cdot\left(\nabla\cdot\bm\sigma\right)=\nabla\cdot\left(\bm u\cdot\bm\sigma\right)-\bm\sigma:\nabla\bm u$, which holds pointwise inside each microphase, where the flow velocity is continuous. Inside each microphase, we thus obtain
	\begin{equation}
		\bm u^*\cdot\bm{\mathcal F}-\bm u\cdot\bm f^*=\nabla\cdot\left(\bm u\cdot\bm\sigma_\text{d}^*-\bm u^*\cdot\bm\sigma_\text{d}\right)+\bm\sigma_\text{d}:\nabla\bm u^*-\bm\sigma_\text{d}^*:\nabla\bm u.\label{eq:LRT-local}
	\end{equation}
	The Lorentz reciprocal theorem relies on the symmetry of the last two terms in this relation. Using Eq. (\ref{eq:-23}) and the constitutive form of the hydrodynamic stress in both flow problems, we have $\bm\sigma_\text{d}:\nabla\bm u^*=\bm\Sigma_\text{m}:\nabla\bm u^*+2\eta\bm e:\bm e^*$. As the dynamic viscosity $\eta$ is the same in both flow problems under kinematic reversibility, the last term on the right-hand side is symmetric in the interchange of $\bm e$ and $\bm e^*$. We can therefore write
	\begin{equation}
		\bm\sigma_\text{d}:\nabla\bm u^*-\bm\sigma_\text{d}^*:\nabla\bm u=\bm\Sigma_\text{m}:\nabla\bm u^*-\bm\Sigma_\text{m}^*:\nabla\bm u.\label{eq:LRT-Sigma}
	\end{equation}
	The right-hand side of Eq. (\ref{eq:LRT-Sigma}) only refers to the microparticles, where the constraint stresses $\bm\Sigma_\text{m}$ and $\bm\Sigma_\text{m}^*$ are defined. Since these stresses are symmetric, only the symmetric parts of the velocity gradients contribute, such that $\bm\Sigma_\text{m}:\nabla\bm u^*=\bm\Sigma_\text{m}:\bm e^*$ and $\bm\Sigma_\text{m}^*:\nabla\bm u=\bm\Sigma_\text{m}^*:\bm e$. As both strain rate tensors vanish inside the rigid microparticles, the right-hand side of Eq. (\ref{eq:LRT-Sigma}) vanishes pointwise, irrespective of the spatial dependence of the constraint stresses and irrespective of whether the microparticles rotate. Inside each microphase, Eq. (\ref{eq:LRT-local}) thus reduces to
	\begin{equation}
		\bm u^*\cdot\bm{\mathcal F}-\bm u\cdot\bm f^*=\nabla\cdot\left(\bm u\cdot\bm\sigma_\text{d}^*-\bm u^*\cdot\bm\sigma_\text{d}\right).\label{eq:-69}
	\end{equation}
	
	In contrast to previous treatments,\cite{happel2012low, teubner1982motion, brady2021phoretic, ganguly2024unified} where Eq. (\ref{eq:-69}) is integrated over the partial volume of the fluid only, we integrate it over the entire region $\mathcal V$ of the RVE, giving
	\begin{eqnarray}
		\underbrace{\int_{\mathcal V}\bm u^*\cdot\bm{\mathcal F}\,d^3\bm r}_{I_1}&-&\underbrace{\int_{\mathcal V}\bm u\cdot\bm f^*\,d^3\bm r}_{I_2}\nonumber\\
        &=&\underbrace{\int_{\mathcal V}\nabla\cdot\left(
			\bm u\cdot\bm\sigma_\text{d}^*-\bm u^*\cdot\bm\sigma_\text{d}
			\right)\,d^3\bm r}_{I_3},\nonumber\\
            \label{eq:-22}
	\end{eqnarray}
	where, in accordance with Eq. (\ref{eq:-69}), the divergence in $I_3$ is evaluated inside each microphase. Based on the relations for the auxiliary flow problem, integrals $I_1$ and $I_2$ can be re-expressed as
	\begin{eqnarray}
		&&I_1=\bm v^*\cdot\int_{\mathcal V}\bm U^{*\dagger}\cdot\bm{\mathcal F}\,d^3\bm r,\label{eq:-90}\\
		&&I_2=N_c\bm v^*\cdot\bm\xi\cdot\bm v.\label{eq:-91}
	\end{eqnarray}
	When Gauss' theorem is applied to integral $I_3$ inside each microphase, the contributions from the two sides of every internal boundary combine into a surface integral of the jump of $\left(\bm u\cdot\bm\sigma_\text{d}^*-\bm u^*\cdot\bm\sigma_\text{d}\right)\cdot\bm n$ across that boundary. Since the hydrodynamic tractions $\bm\sigma_\text{d}\cdot\bm n$ and $\bm\sigma_\text{d}^*\cdot\bm n$ are continuous across all boundaries (Section \ref{sec:-4}), and since the normal component of the flow velocity is continuous across the impermeable surfaces of the microparticles, this jump only stems from a jump in the tangential flow velocity, giving
	\begin{eqnarray}
		I_3&=&\oint_{\partial\mathcal V}\left(\bm u\cdot\bm\sigma_\text{d}^*-\bm u^*\cdot\bm\sigma_\text{d}\right)\cdot\,\bm n\,dS\nonumber\\
        &&-\oint_{\partial\nu_\text{m}}\left(\bm u^{\parallel}\cdot\bm\sigma_\text{d}^*-\bm u^{\parallel *}\cdot\bm\sigma_\text{d}\right)\cdot\,\bm n\,dS,\label{eq:-17}
	\end{eqnarray}
	where $\bm u^{\parallel}$ and $\bm u^{\parallel *}$ are the tangential slip velocities at the surfaces $\partial\nu_\text{m}$ of the microparticles, defined as the jump of the tangential flow velocity from the microparticle surface to the adjacent solution in each flow problem. The surface integral over $\partial\nu_\text{m}$ in Eq. (\ref{eq:-17}) thus collects all interfacial contributions to $I_3$, and vanishes identically for a no-slip boundary condition, where the flow velocity is continuous. At boundaries inside the solution, such as the outer boundary of an interfacial layer, both the flow velocity and the hydrodynamic traction are continuous, such that no interfacial contributions arise. Equivalently, Eq. (\ref{eq:-17}) is recovered when all derivatives are treated in the distributional sense (Section \ref{sec:-6}), in which case the jumps in the tangential flow velocity give rise to surface contributions to $\bm\sigma_\text{d}:\nabla\bm u^*-\bm\sigma_\text{d}^*:\nabla\bm u$ that are concentrated on $\partial\nu_\text{m}$ and only involve the continuous hydrodynamic tractions. As thermal transpiration is negligible at low Knudsen number,\cite{Bielenberg2005ACM} these slip velocities satisfy the Navier slip boundary condition:\cite{bedeaux1976boundary}
	\begin{eqnarray}
		&&\bm u^{\parallel}=\bm u_\text{a}+\frac{l_\text{slip}}{\eta}\left(\bm I-\bm n\bm n\right)\cdot\left(\bm n\cdot\bm\sigma_\text{d}\right),\label{eq:-15}\\
		&&\bm u^{\parallel *}=\frac{l_\text{slip}}{\eta}\left(\bm I-\bm n\bm n\right)\cdot\left(\bm n\cdot\bm\sigma_\text{d}^*\right).
	\end{eqnarray}
	In the actual flow problem, the active slip velocity profile $\bm u_\text{a}$ is defined on all microparticle surfaces. Here, this active slip velocity represents a boundary-layer treatment of force-free self-propulsion mechanisms stemming from non-hydrostatic forces that are not described by the hydrostatic force density $\bm{\mathcal F}$, such as those due to ciliary motion.\cite{lighthill1952squirming, blake1971spherical} As for dynamic viscosity, the principle of kinematic reversibility justifies the use of the same hydrodynamic slip length $l_\text{slip}$ in both flow problems. Let us consider the integral of the term $\bm u^{\parallel}\cdot\bm\sigma_\text{d}^*\cdot\bm n$ over the interfaces in Eq. (\ref{eq:-17}). Substituting Eq. (\ref{eq:-15}), we can write $\left[\left(\bm I-\bm n\bm n\right)\cdot\left(\bm n\cdot\bm\sigma_\text{d}\right)\right]\cdot\left(\bm\sigma_\text{d}^*\cdot\bm n\right)=\left(\bm n\cdot\bm\sigma_\text{d}\right)\cdot\left(\bm n\cdot\bm\sigma_\text{d}^*\right)-\left(\bm n\cdot\bm\sigma_\text{d}\cdot\bm n\right)\left(\bm n\cdot\bm\sigma_\text{d}^*\cdot\bm n\right)$, which is symmetric in the interchange of $\bm\sigma_\text{d}$ and $\bm\sigma_\text{d}^*$. Hence, we obtain
	\begin{equation}
		\oint_{\partial\nu_\text{m}}\left(\bm u^{\parallel}\cdot\bm\sigma_\text{d}^*-\bm u^{\parallel *}\cdot\bm\sigma_\text{d}\right)\cdot\,\bm n\,dS=\bm v^*\cdot\oint_{\partial\nu_\text{m}}\bm T^{*\dagger}\cdot\bm u_\text{a}\,dS.\label{eq:-61}
	\end{equation}
	
    To describe colloidal diffusion, we now introduce the decompositions $\bm u=\bm u'+\left\langle\bm u\right\rangle$ and $\bm \sigma_\text{d}=\bm\sigma_\text{d}'+\left\langle\bm \sigma_\text{d}\right\rangle$ into the surface integral over $\partial\mathcal V$ in Eq. (\ref{eq:-17}). In the dilute limit, the flow velocities $\bm u'$ and $\bm u^*$ and hydrodynamic stress tensors $\bm \sigma_\text{d}'$ and $\bm \sigma_\text{d}^*$ are known to decay as $1/r$ and $1/r^2$ inside the solution, respectively,\cite{happel2012low, batchelor1970stress} implying that the corresponding surface integrals over $\partial\mathcal V$ vanish: 
    \begin{equation}
		\oint_{\partial\mathcal V}\left(\bm u'\cdot\bm\sigma_\text{d}^*-\bm u^*\cdot\bm\sigma_\text{d}'\right)\cdot\,\bm n\,dS=0.\label{eq:-58}
	\end{equation}
    Notably, Eqs. (\ref{eq:-61}) and (\ref{eq:-58}) specifically rely on the properties of hydrodynamic stress induced by diffusion flow, thereby highlighting the necessity of decoupling this stress from the hydrostatic stress. For a homogeneous, isotropic suspension beyond the dilute limit, Eq. (\ref{eq:-58}) remains valid due to the statistical invariance of these fields under translation, which amounts to neglecting finite-size fluctuations of this surface integral for a sufficiently large RVE, in accordance with the thermodynamic-limit approximation introduced in Section \ref{sec:-5}. Moreover, Eqs. (\ref{eq:avg-grad}) and (\ref{eq:-16}), applied to each component of $\bm u^*$, give $\oint_{\partial\mathcal V}\bm n\otimes\bm u^*\,dS=V\bar\nabla\left\langle\bm u^*\right\rangle=0$, such that $\oint_{\partial\mathcal V}\bm u^*\cdot\left\langle\bm\sigma_\text{d}\right\rangle\cdot\bm n\,dS=0$, whereas the macroscopic momentum balance of the auxiliary problem gives $\oint_{\partial\mathcal V}\bm\sigma_\text{d}^*\cdot\bm n\,dS=-N_c\bm F^*$. As a result, the surface integral over $\partial\mathcal V$ in Eq. (\ref{eq:-17}) reduces to
	\begin{equation}
		\oint_{\partial\mathcal V}\left(\bm u\cdot\bm\sigma_\text{d}^*-\bm u^*\cdot\bm\sigma_\text{d}\right)\cdot\,\bm n\,dS=-\bm v^*\cdot N_c\bm\xi\cdot\left\langle\bm u\right\rangle.\label{eq:-63}
	\end{equation}
	Using Eqs. (\ref{eq:-61}) and (\ref{eq:-63}) in Eq. (\ref{eq:-17}), integral $I_3$ thus takes the form
	\begin{equation}
		I_3=-\bm v^*\cdot N_c\bm\xi\cdot\left\langle\bm u\right\rangle-\bm v^*\cdot\oint_{\partial\nu_\text{m}}\bm T^{*\dagger}\cdot\bm u_\text{a}\,dS.\label{eq:-20}
	\end{equation}
	Substituting Eqs. (\ref{eq:-90}), (\ref{eq:-91}) and (\ref{eq:-20}) into Eq. (\ref{eq:-22}), the following relation is obtained:
    \begin{eqnarray}
		&&\bm v^*\cdot N_c\bm\xi\cdot\left(\bm v-\left\langle\bm u\right\rangle\right)\\
        &&=\bm v^*\cdot\left(\int_{\mathcal V}\bm U^{*\dagger}\cdot\bm{\mathcal F}\,d^3\bm r+\oint_{\partial\nu_\text{m}}\bm T^{*\dagger}\cdot\bm u_\text{a}\,dS\right).
	\end{eqnarray}
    Since this relation holds for any sedimentation velocity $\bm v^*$, the colloidal velocity relative to advection, or colloidal diffusion velocity, can finally be identified as	
	\begin{equation}
		\bm v-\left\langle\bm u\right\rangle=\frac{\bm\xi^{-1}}{N_c}\cdot\left(\int_{\mathcal V}\bm U^{*\dagger}\cdot\bm{\mathcal F}\,d^3\bm r+\oint_{\partial\nu_\text{m}}\bm T^{*\dagger}\cdot\bm u_\text{a}\,dS\right).\label{eq:-94}
	\end{equation}
	The corresponding colloidal diffusion flux $\bm{\mathcal J}_c$ is then obtained by multiplying Eq. (\ref{eq:-94}) by the colloidal number density $n_c$, yielding:
    \begin{equation}
		\bm{\mathcal J}_c=\bm\xi^{-1}\cdot\left(\left\langle\bm U^{*\dagger}\cdot\bm{\mathcal F}\right\rangle+\frac{S_\text{m}}{V}\left\langle\bm T^{*\dagger}\cdot\bm u_\text{a}\right\rangle_{\partial\nu_\text{m}}\right),\label{eq:-84}
	\end{equation}
    where $S_\text{m}$ is the total interfacial surface area of the microparticles inside $\mathcal V$ and where $\langle\cdot\rangle_{\partial\nu_\text{m}}$ is the corresponding surface average.
    
    Crucially, the colloidal diffusion velocity $\bm v-\left\langle\bm u\right\rangle$ given by Eq. (\ref{eq:-94}) only refers to the hydrodynamic tensors $\bm U^*$ and $\bm T^*$ of the auxiliary Stokes problem of colloidal sedimentation, which is well established and feasible to study numerically by applying a bulk flow within an RVE containing an ensemble of stationary colloids. In the fluid region, these tensors account for the hydrodynamic drag on the colloids due to forces on the fluid components. So-called "thermodiffusive" approaches\cite{PhysRevLett.95.208301,dhont2007thermodiffusion} ignore this drag by assuming that the colloid's hydrodynamic radius tends to zero, which corresponds to the Hückel limit of colloidal diffusion.\cite{morthomas2008thermoelectric} The first term in Eq. (\ref{eq:-94}) describes diffusion due to vectorial thermodynamic forces, including sedimentation and phoretic transport in response to externally applied or self-generated thermodynamic forces. The latter forces are caused by chemical or thermal activity, whereas the forces giving rise to the second term in Eq. (\ref{eq:-94}) stem from ciliary activity. In this context, it should be noted that any form of activity eventually produces correlations in colloidal motion that result in a breakdown of homogeneity and isotropy inside the suspension.\cite{liebchen2015clustering}
	
	To demonstrate the connection to Onsager's reciprocal theory, we now substitute Eq. (\ref{eq:-88}) into Eq. (\ref{eq:-84}), yielding 
	\begin{eqnarray}
		\bm{\mathcal J}_c&=&-\bm\xi^{-1}\cdot\left\langle\check h^\text{eq}\bm U^{*\dagger}\cdot\frac{\nabla T}{T^\text{eq}}\right\rangle\nonumber\\
        &&+\bm\xi^{-1}\cdot\sum_i\left\langle n_i^\text{eq}\bm U^{*\dagger}\cdot\left(\bm F_i-\nabla_{T,P_\text{d}}\mathring\mu_i\right)\right\rangle\nonumber\\
        &&+\bm\xi^{-1}\cdot\frac{S_\text{m}}{V}\left\langle\bm T^{*\dagger}\cdot\bm u_\text{a}\right\rangle_{\partial\nu_\text{m}}.
	\end{eqnarray}
	Indeed, the quantities $\check h^\text{eq}\bm U^*$ and $n_i^\text{eq}\bm U^*$ can be identified as the local flows of heat and component particles induced by colloidal sedimentation, respectively. If the thermodynamic forces are uniform over $\mathcal V$, which amounts to ignoring interfacial boundary conditions inside $\mathcal V$, we obtain 
	\begin{eqnarray}
		\bm{\mathcal J}_c&=&-\bm L_{c q}\cdot\frac{\nabla T}{T^\text{eq}}+\sum_i\bm L_{c i}\cdot\left(\bm F_i-\nabla_{T,P_\text{d}}\mathring\mu_i\right)\nonumber\\
        &&+\bm\xi^{-1}\cdot\frac{S_\text{m}}{V}\left\langle\bm T^{*\dagger}\cdot\bm u_\text{a}\right\rangle_{\partial\nu_\text{m}},
	\end{eqnarray}
	with $\bm L_{c q}=\bm\xi^{-1}\cdot\left\langle\check h^\text{eq}\bm U^{*\dagger}\right\rangle$ and $\bm L_{c i}=\bm\xi^{-1}\cdot\left\langle n_i^\text{eq}\bm U^{*\dagger}\right\rangle$. Since transport tensors reduce to scalars in a homogeneous, isotropic suspension, such that $\bm L_{c q}=L_{c q}\bm I$, $\bm L_{c i}=L_{c i}\bm I$ and $\bm \xi=\xi\bm I$, we finally recover the standard scalar form of the Onsager reciprocal relations for colloidal diffusion: 
	\begin{eqnarray}
		&&L_{c q}\bm I=L_{qc}\bm I=\frac{1}{\xi}\left\langle\check h^\text{eq}\bm U^{*\dagger}\right\rangle,\label{eq:-104}\\
		&&L_{c i}\bm I=L_{ic}\bm I=\frac{1}{\xi}\left\langle n_i^\text{eq}\bm U^{*\dagger}\right\rangle\label{eq:-105}.
	\end{eqnarray}
	The Hückel limit, for which $\bm U^* = 0$ in the fluid region, thus corresponds to the case of vanishing hydrodynamic cross-coupling between the colloidal diffusion flux and the thermodynamic forces in the fluid. Based on the assumption of uniform gradients, it can also be concluded that the standard scalar form of the Onsager reciprocal relations cannot account for active motion.
	
	To obtain analytical expressions for the auxiliary hydrodynamic tensors, we may consider the dilute limit where the diffusion of a single colloid can be studied inside an unbounded solution. Assuming uniform viscosity ($\eta=\eta^\text{b}$), the friction tensor of an isolated, spherical, homogeneous colloid of hydrodynamic radius $R$ is given by
	\begin{equation}
		\bm\xi=\xi\bm I=6\pi a\eta R\bm I,\label{eq:-73}
	\end{equation}
	where the hydrodynamic slip parameter $a=(1+2l_\text{slip}/R)/(1+3l_\text{slip}/R)$ takes the value $1$ for a no-slip boundary condition and $2/3$ for a perfect-slip boundary condition. To leading order, the backflow required to satisfy Eq. (\ref{eq:-16}) scales as $1/L$, where $L$ is the macroscopic length scale. In an unbounded solution, the auxiliary hydrodynamic tensors therefore take the familiar symmetric forms:
	\begin{eqnarray}
		&&\bm U^*=\frac{3R}{4r}\left[a\left(\bm I+\hat{\bm r}\hat{\bm r}\right)-\left(a-\frac{2}{3}\right)\frac{R^2}{r^2}\left(3\hat{\bm r}\hat{\bm r}-\bm I\right)\right],\label{eq:-5}\\
		&&\bm T^*=-\frac{9\eta R}{2r^2}
		\left[a\hat{\bm r}\hat{\bm r}-\left(a-\frac{2}{3}\right)\frac{R^2}{r^2}\left(3\hat{\bm r}\hat{\bm r}-\bm{I}\right)\right],\label{eq:-75}
	\end{eqnarray}
	where $r$ is the radial distance from the hydrodynamic center of the colloid and $\hat{\bm r}$ is the outward-pointing radial unit vector. Alternatively, Eq. (\ref{eq:-75}) can be expressed as $\bm T^*=9\eta aR\bm{I}/\left(2r^2\right)-6\eta\bm U^*/r$. With Eqs. (\ref{eq:-73}) and (\ref{eq:-75}), the contribution due to ciliary motion in Eq. (\ref{eq:-94}) then reduces to the colloidal swimming velocity $\bm v_\text{swim}=\bm\xi^{-1}\cdot\oint_{\partial\nu_c}\left(\bm T^{*\dagger}\cdot\bm u_\text{a}\right)\,dS$, which becomes
	\begin{equation}
		\bm v_\text{swim}=-\left(3-\frac{2}{a}\right)\frac{1}{4\pi R^2}\oint_{\partial\nu_c}\bm u_\text{a}\,dS=-\frac{\left\langle\bm u_\text{a}\right\rangle_{\partial\nu_c}}{1+2\frac{l_\text{slip}}{R}}.\label{eq:-42}
	\end{equation}
	This expression vanishes for a perfect-slip boundary condition ($a=2/3$), whereas it reduces to the familiar form $-\left(4\pi R^2\right)^{-1}\oint_{\partial\nu_c}\bm u_\text{a}\,dS$ for a no-slip boundary condition ($a=1$),\cite{stone1996propulsion} thereby generalizing the latter result to arbitrary hydrodynamic surface slippage at the colloidal surface. A vanishing swimming velocity for $a=2/3$ is indeed expected, since a perfectly slipping ciliary envelope cannot exert a tangential force on the fluid.
	
	\subsection{Fluid-mechanical formulation of colloidal diffusion}\label{sec:-1}
	
	A fluid-mechanical formulation of colloidal diffusion is based on the premise that the equations of state are known for the microphases of the fluid, but unknown for the colloidal microphase. Consequently, the hydrostatic forces acting within the colloids must be eliminated from Eq. (\ref{eq:-94}) to obtain a computable, fluid-mechanical form of the colloidal diffusion flux. 
    
\begin{figure}
		\centering
		\includegraphics[width=0.3\textwidth]{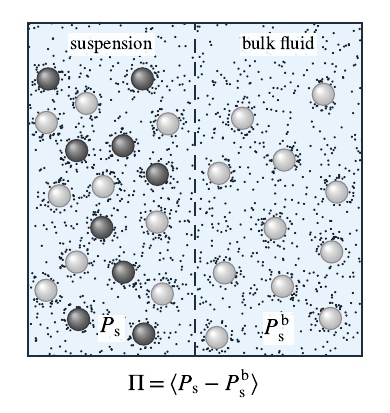}
		\caption{From a continuum-mechanical perspective, the osmotic pressure of the colloids (dark spheres) is defined as the pressure difference across a semi-permeable membrane (dashed line) separating the bulk fluid from the suspension.}
		\label{fig:-4}
	\end{figure}

    To this end, we consider the volume integral of $\bm{\mathcal F}$ over $\mathcal V$ and write it as a sum of integrals over the colloidal region $\nu_c$ and fluid region $\tilde\nu$, which yields $\int_{\nu_c}\bm{\mathcal F}\,d^3\bm r+\int_{\tilde\nu}\bm{\mathcal F}\,d^3\bm r=\int_{\mathcal V}\rho\,d^3\bm r\,\bm g-\int_{\mathcal V}\nabla P_\text{s}\,d^3\bm r$ due to electroneutrality of $\mathcal V$ (Section \ref{sec:-4}). Inside each colloid, the auxiliary flow velocity corresponds to a rigid-body motion, whose volume average over the colloidal microphase is given by Eq. (\ref{eq:-18}). If the colloids translate uniformly without rotating in the auxiliary problem, which is exactly the case for an isolated spherical colloid, we have $\bm U^*=\bm I$ everywhere inside $\nu_c$, and the volume integral in Eq. (\ref{eq:-94}) can be written as
	\begin{equation}
		\int_{\mathcal V}\bm U^{*\dagger}\cdot\bm{\mathcal F}\,d^3\bm r=\int_{\nu_c}\bm{\mathcal F}\,d^3\bm r+\int_{\tilde\nu}\bm U^{*\dagger}\cdot\bm{\mathcal F}\,d^3\bm r.\label{eq:Uc-split}
	\end{equation}
	More generally, Eq. (\ref{eq:Uc-split}) neglects the correlation $\int_{\nu_c}\left(\bm U^{*\dagger}-\bm I\right)\cdot\bm{\mathcal F}\,d^3\bm r$, which couples the net hydrostatic force and torque acting on each colloid to the deviation of its auxiliary translational and rotational velocity from the mean sedimentation velocity $\bm v^*$. This correlation vanishes for an isolated colloid and whenever all colloids experience the same net force without torque. In particular, it does not affect the contribution of the colloidal osmotic pressure gradient, since the thermal and interparticle forces represented by this gradient are equivalent to a uniform force on each colloid in a locally equilibrated suspension,\cite{batchelor1976brownian} which leads to the generalized Einstein relation given at the end of this section. A nonzero correlation can, however, arise from first-order forces that differ from one colloid to another, for instance when neighboring microparticles distort the local thermodynamic forces and electric field that drive phoretic motion, or when the colloids experience a torque. Its neglect beyond the dilute limit constitutes an approximation, which is, however, not required for the reciprocal form of the colloidal diffusion velocity given by Eq. (\ref{eq:-94}). Combining these two relations, the net hydrostatic driving force of colloidal diffusion, given by the first term in Eq. (\ref{eq:-94}), becomes
	\begin{eqnarray}
		\int_{\mathcal V}\bm U^{*\dagger}\cdot\bm{\mathcal F}\,d^3\bm r&=&\int_{\tilde\nu}\left(\bm U^{*\dagger}-\bm I\right)\cdot\bm{\mathcal F}\,d^3\bm r\nonumber\\
        &&+\int_{\mathcal V}\rho\,d^3\bm r\,\bm g-\int_{\mathcal V}\nabla P_\text{s}\,d^3\bm r.\label{eq:-96}
	\end{eqnarray}
    In the fluid region, the hydrostatic force density $\bm{\mathcal F}$
    can be decomposed into a bulk contribution $\bm{\mathcal F}^\text{b}$ and a fluid-mechanical disturbance $\delta\bm{\mathcal F}$, such that $\bm{\mathcal F}=\delta\bm{\mathcal F}+\bm{\mathcal F}^\text{b}$.
    Since the bulk fluid is defined in the absence of the colloids, its hydrostatic force density $\bm{\mathcal F}^\text{b}$ has no statistical correlation with the Stokes flow tensor $\bm U^*$, and we further have 
    \begin{equation}
        \int_{\tilde\nu}\left(\bm U^{*\dagger}-\bm I\right)\cdot\bm{\mathcal F}\,d^3\bm r=\int_{\tilde\nu}\left(\bm U^{*\dagger}-\bm I\right)\cdot\delta\bm{\mathcal F}\,d^3\bm r-\int_{\mathcal V}\bm{\mathcal F}^\text{b}\,d^3\bm r,\label{eq:-66}
    \end{equation}
    where Eqs. (\ref{eq:-16}) and (\ref{eq:-18}) were also used. 
    Although the bulk fluid may not be electroneutral everywhere within $\mathcal V$ if it contains other species of microparticles ($\omega^\text{b}\ne 0$), it must remain overall electroneutral over $\mathcal V$, giving
    \begin{equation}
        \int_{\mathcal V}\bm{\mathcal F}^\text{b}\,d^3\bm r=\int_{\mathcal V}\rho^\text{b}\,d^3\bm r\,\bm g-\int_{\mathcal V}\nabla P_\text{s}^\text{b}\,d^3\bm r,\label{eq:-154}
    \end{equation}
    where $P_\text{s}^\text{b}$ is the hydrostatic pressure of the bulk fluid. As illustrated in Fig. \ref{fig:-4}, the colloidal osmotic pressure $\Pi$ can further be introduced via $\Pi=\left\langle P_\text{s}-P_\text{s}^\text{b}\right\rangle=\left\langle\delta P_\text{s}\right\rangle$. With gradients understood in the distributional sense (Section \ref{sec:-6}), Eq. (\ref{eq:avg-grad}) then gives
	\begin{equation}
		\bar\nabla\Pi=\left\langle\nabla\delta P_\text{s}\right\rangle,\label{eq:-147}
	\end{equation}
	where $\nabla\delta P_\text{s}$ includes the surface contributions arising from jumps in the hydrostatic pressure across boundaries, consistent with the treatment of the integral of $\nabla P_\text{s}$ in Eq. (\ref{eq:-96}).

    Substituting Eqs. (\ref{eq:-96})-(\ref{eq:-154}) into Eq. (\ref{eq:-94}), and using Eq. (\ref{eq:-147}), the colloidal diffusion velocity finally takes the fluid-mechanical form
	\begin{eqnarray}
		\bm v-\left\langle\bm u\right\rangle&=&\frac{\bm\xi^{-1}}{N_c}\cdot\left[\int_{\tilde\nu}\left(\bm U^{*\dagger}-\bm I\right)\cdot\bm{\mathcal F}_\text{osm}\,d^3\bm r\right.\nonumber\\
		&&\hspace{1.2cm}+\left(\int_{\nu_c}\delta\rho\bm I\,d^3\bm r+\int_{\tilde\nu}\delta\rho\bm U^{*\dagger}\,d^3\bm r\right)\cdot\bm g\nonumber\\
		&&\hspace{1.2cm}\left.-V\bar\nabla\Pi+\oint_{\partial\nu_\text{m}}\bm T^{*\dagger}\cdot\bm u_\text{a}\,dS\right],\label{eq:-100}
	\end{eqnarray}
	where $\bm U^*-\bm I$ represents the fluid flow velocity induced by sedimentation in the rest frame of the colloids. As in Eq. (\ref{eq:-94}), the last term in Eq. (\ref{eq:-100}) accounts for ciliary motion. The osmotic force density $\bm{\mathcal F}_\text{osm}$ is defined by
	\begin{equation}
        \bm{\mathcal F}_\text{osm}=\delta\left[\omega\bm E-\left(\check h^\text{eq}\frac{\nabla T}{T^\text{eq}}+\sum_i n_i^\text{eq}\nabla_{T,P_\text{d}}\mathring\mu_i\right)\right].\label{eq:-102}
	\end{equation}
    It is related to the fluid-mechanical disturbance of the hydrostatic force density via $\delta\bm{\mathcal F}=\bm{\mathcal F}_\text{osm}+\delta\rho\,\bm g$, since $\sum_i n_i^\text{eq}\bm F_i=\omega\bm E+\rho\bm g$. Inside the solution, Eq. (\ref{eq:-102}) can equivalently be evaluated using the reduced form of the hydrostatic force density given by Eq. (\ref{eq:-67}), since the part $\bm{\mathcal F}\left(\left\{\phi_j\right\}=0\right)$ is balanced by the solvent both in the presence and in the absence of the colloids, such that its fluid-mechanical disturbance vanishes. Evidently, the fluid-mechanical disturbance of the electric force density $\omega\bm E$ in Eq. (\ref{eq:-102}) drives electrophoresis, whereas the remaining disturbance drives diffusio- and thermophoresis. Based on Eq. (\ref{eq:-65}), the latter contribution can be identified as the local hydrostatic pressure gradient evaluated at constant interfacial interaction potentials, such that 
    \begin{equation}
		\nabla_{\left\{\phi_j\right\}} P_\text{s}=\check h^\text{eq}\frac{\nabla T}{T^\text{eq}}+\sum_i n_i^\text{eq}\nabla_{T,P_\text{d}}\mathring\mu_i.\label{eq:-64}
	\end{equation}
     Alternatively, Eq. (\ref{eq:-102}) can therefore be written as
    \begin{equation}
        \bm{\mathcal F}_\text{osm}=\delta\left(\omega\bm E\right)-\nabla_{\left\{\phi_j\right\}}\delta P_\text{s}.\label{eq:-121}
	\end{equation}
	Based on overall electroneutrality of $\mathcal V$, and noting that interfacial potential interactions cause equal and opposite forces on the microphases inside $\mathcal V$, it also directly follows that
    \begin{equation}
        \int_{\mathcal V}\bm{\mathcal F}_\text{osm}\,d^3\bm r=-V\bar\nabla\Pi.\label{eq:Fosm-int}
    \end{equation}
    The osmotic force density is therefore overall force-free when the colloidal osmotic pressure is uniform, which corresponds to the well-known force-free condition of phoretic motion.\cite{anderson1989colloid}
	
	To first order in the non-equilibrium forces, Eq. (\ref{eq:-100}) provides a complete description of colloidal diffusion in a homogeneous suspension and reduces to the diffusion velocity of a single colloid in the dilute limit. The first term represents (self-)phoretic motion, which is consistent with its force-free nature: overall, fluid flows in the direction of the osmotic force density $\bm{\mathcal F}_\text{osm}$, whereas the colloid is pushed in the opposite direction. In the dilute limit, the BLA for phoretic motion of a single colloid, applied to thermophoresis by Derjaguin\cite{churaev2013surface} and known as the Smoluchowski limit in electrophoresis, can be recovered from this term by assuming uniform gradients and a no-slip hydrodynamic boundary condition. Using Eq. (\ref{eq:-5}) and expanding this term to first order in $z=r-R\ll R$, one then recovers the well-known BLA expression\cite{burelbach2018unified}
	\begin{equation}
		\int_{\tilde\nu}\left(\bm U^{*\dagger}-\bm I\right)\cdot\bm{\mathcal F}_\text{osm}\,d^3\bm r\approx -4\pi R\int_0^\infty z\bm{\mathcal F}_\text{osm}\,dz.
	\end{equation} 
    
    The second term in Eq. (\ref{eq:-100}) represents buoyant sedimentation, accounting for the buoyant mass $\int_{\nu_c}\delta\rho\,d^3\bm r$ of the colloids and for the hydrodynamic drag caused by the equilibrium-interfacial excess weight of fluid via $\int_{\tilde\nu}\delta\rho\bm U^*\,d^3\bm r$. The third term has a purely continuum-statistical nature, accounting for the average colloidal motion due to thermal fluctuations and hydrostatic many-body interactions via the colloidal osmotic pressure gradient $\bar\nabla\Pi$.\cite{dhont2004thermodiffusion} When a colloidal concentration gradient is imposed at uniform temperature, the corresponding contribution $-\bm\xi^{-1}\cdot\bar\nabla\Pi$ to the colloidal diffusion flux can indeed be written as $-\bm D\cdot\bar\nabla n_c$, where the diffusion tensor $\bm D$ satisfies the generalized Einstein relation 
    \begin{equation}
        \bm\xi\cdot\bm D=\frac{\partial\Pi}{\partial n_c}\bm I.
    \end{equation}
    Notably, all terms in Eq. (\ref{eq:-100}) have been derived from a single application of the Lorentz reciprocal theorem using the auxiliary Stokes flow problem of sedimentation.

    \section{Discussion and application to diffusiophoresis resulting from an interfacial step potential}\label{sec:-2}
    
    \subsection{Comparison with previous approaches}\label{sec:IVA}
    
    In the absence of ciliary motion, Eq. (\ref{eq:-100}) aligns well with a previously developed non-equilibrium thermodynamic approach based on Onsager reciprocity,\cite{burelbach2018unified} which has been verified numerically using mesoscale-molecular simulations,\cite{Burelbach2018ThermophoreticFO} applied successfully to experimental data on thermophoresis of DNA,\cite{Burelbach2018DeterminingPM} and used to generalize the Henry function for electrophoresis and the Ruckenstein term for thermophoresis to arbitrary hydrodynamic surface slippage and interfacial layer thicknesses.\cite{burelbach2019particle} However, when applying this approach to colloidal diffusion due to non-uniform thermodynamic forces, the author assumed that the fluid-mechanical disturbance of the conservative part of the hydrostatic force density can drive flow, despite the fact that the solvent balances precisely this part. As a result, it was incorrectly concluded that phoretic motion of a single colloid may occur in the absence of interfacial potential interactions.\cite{burelbach2019particle,burelbach2019linear} The interfacially driven predictions of these treatments are not affected by this issue, as shown in Appendix \ref{app:validity}. The continuum-statistical framework presented here resolves this issue and demonstrates that reciprocity in colloidal motion emerges from kinematic reversibility under the assumption of RTE.
    
    Eq. (\ref{eq:-100}) can also be compared with more recent treatments of phoretic motion that go beyond the BLA. For a single colloid suspended in an unbounded solution, the phoretic contribution in Eq. (\ref{eq:-100}) has the same structure as the mobility expressions derived by Brady and by Ganguly \textit{et al.} from the Lorentz reciprocal theorem for a colloid in an unbounded bath,\cite{brady2021phoretic,ganguly2024unified} in which the disturbance velocity $\bm U^*-\bm I$ of the auxiliary Stokes problem weights a body force density exerted on the fluid. In these treatments, the body force density is expressed in terms of the interaction force between the solute and the colloid, in some cases supplemented by an osmotic contribution,\cite{ganguly2024unified} whereas Eq. (\ref{eq:-100}) involves the osmotic force density $\bm{\mathcal F}_\text{osm}$, defined as a fluid-mechanical disturbance of the hydrostatic force density. These forms differ at most by gradient terms. For an ideal solute, denoted by an index value $i=1$, for instance, the body force density exerted on the fluid can be written as $-n_1\nabla\phi_1=\bm{\mathcal F}_\text{osm}+\nabla\Psi_1$, where $\Psi_1=n_1k_BT-n_1^{\text{b}|\text{eq}}\mathring\mu_1$ corresponds, up to a uniform contribution, to the excess osmotic pressure of the solute relative to an ideal solution in local equilibrium with $\mathring\mu_1$ in the absence of the interfacial interaction, and where $k_B$ is the Boltzmann constant. When this excess osmotic contribution is included, as in Ref.~\onlinecite{ganguly2024unified}, the body force density coincides with $\bm{\mathcal F}_\text{osm}$. For the interaction force alone, the gradient term $\nabla\Psi_1$ vanishes outside the interfacial layer to first order. Such terms do not contribute to the phoretic velocity: for any scalar field $\Psi$ whose deviation from a uniform value decays faster than $1/r^2$ far from the colloid, Gauss' theorem gives
    \begin{equation}
    	\int_{\tilde\nu}\left(\bm U^{*\dagger}-\bm I\right)\cdot\nabla\Psi\,d^3\bm r=0,\label{eq:grad-zero}
    \end{equation}
    since $\nabla\cdot\bm U^*=0$ and the normal component of $\bm U^*-\bm I$ vanishes at the colloidal surface, irrespective of the hydrodynamic slip. Using a different route, Marbach \textit{et al.} obtained the diffusiophoretic velocity of a no-slip sphere for an arbitrary interaction range by solving the Stokes equations directly and imposing the force-free condition.\cite{marbach2020local} Their result has the same structure as the phoretic contribution in Eq. (\ref{eq:-100}): the radial weighting function in their expression coincides with the angular average of $\bm U^*-\bm I$ for a no-slip sphere, and their force density, given by the interaction force, differs from $\bm{\mathcal F}_\text{osm}$ only by a gradient term. For a single colloid, the present approach therefore recovers these previous treatments, but it adds value in several respects. First, Eq. (\ref{eq:-100}) results from a reciprocal formulation at the scale of an RVE of a homogeneous suspension, such that phoretic motion, buoyant sedimentation and the colloidal osmotic pressure gradient emerge from a single application of the Lorentz reciprocal theorem, together with the Onsager reciprocal relations of colloidal transport. Second, the osmotic force density is expressed in terms of thermodynamic forces and equilibrium-interfacial excess densities, which places diffusio-, thermo- and electrophoresis on the same footing without requiring the interaction force itself, and which separates the contributions that drive phoretic motion from those that are balanced by the solvent. Third, the structure of the phoretic contribution has been shown to hold for a general Navier slip boundary condition at the colloidal surface, whereas the treatments above assumed a no-slip boundary condition.
    
    \subsection{Application to an interfacial step potential}\label{sec:IVB}
    
    \begin{figure}
    	\centering
    	\includegraphics[width=0.4\textwidth]{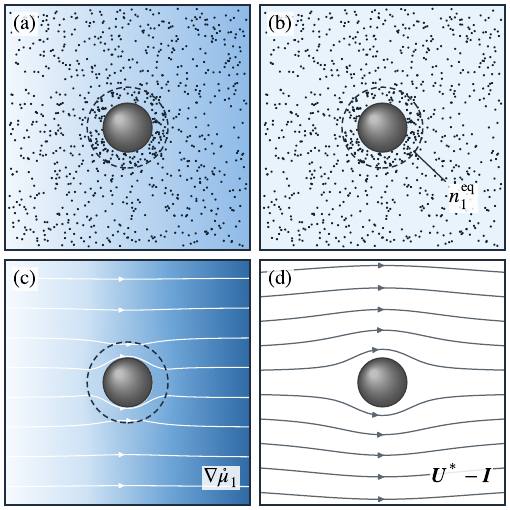}
    	\caption{Schematic representation of the reciprocal approach to colloidal motion. a) A single colloid subjected to non-equilibrium forces. The continuum fields giving rise to these forces are illustrated by a blue background gradient. The dashed circle around the colloid represents the effective boundary between its interfacial region and the bulk fluid. Instead of resolving the full local dynamics of a), the reciprocal approach consists of three separate parts: b) An equilibrium-statistical problem to determine local component distributions, and a continuum-mechanical problem to independently compute c) the non-equilibrium forces and d) the induced Stokes flow field. A stronger background gradient is used in c) to show that the colloid can modify the continuum fields in its vicinity if the transport properties of its solid microphase differ from those of the solution.}
    	\label{fig:-2}
    \end{figure}
    
    As summarized in Fig. \ref{fig:-2} for a single colloid, Eq. (\ref{eq:-100}) reduces the computation of the colloidal diffusion velocity to three independent problems: an equilibrium-statistical problem to determine the equation of state and the equilibrium component distributions of the fluid (Fig. \ref{fig:-2}b), a steady-state transport problem to determine the local thermodynamic forces under the given non-equilibrium boundary conditions (Fig. \ref{fig:-2}c), and a hydrodynamic problem to compute the auxiliary Stokes flow of sedimentation under stationary boundary conditions (Fig. \ref{fig:-2}d). 
    
    To illustrate this procedure, and to validate the framework against established results, we now apply it to the diffusiophoresis of a single colloid inside an unbounded solution. Assuming that the system is at uniform temperature, we replace the notation $\nabla_T$ by $\nabla$ for chemical potential gradients. The colloid, whose solid boundary is chosen to coincide with a hydrodynamic boundary of radius $R$, is suspended in a solution composed of a single solvent component ($i=0$) and a single solute component ($i=1$), such that the fluid region $\tilde\nu$ is represented by the range $r>R$. The solute is assumed to be point-like ($\bar v_1=0$) and to follow an ideal equation of state, such that the dynamic viscosity $\eta$ of the solution is uniform and set by the viscosity of the pure solvent.
    
    \textit{Equilibrium problem (Fig. \ref{fig:-2}b).} The colloid is assumed to interact specifically with the solute only, via a step potential
    \begin{equation}
    	\phi_1=
    	\begin{cases}
    		\mathcal{E}, & R<r\le R+\lambda, \\[1ex]
    		0, & r>R+\lambda,
    	\end{cases}
    \end{equation}
    where $\lambda$ is the thickness of the interfacial layer and $\mathcal{E}$ is the potential energy of a solute particle inside it. The equilibrium solute density then obeys a Boltzmann distribution, with $n_1^\text{eq}=n_1^{\text{b}|\text{eq}}$ outside the layer and $n_1^\text{eq}=Kn_1^{\text{b}|\text{eq}}$ inside it, where 
    \begin{equation}
        K=\exp\left(-\frac{\mathcal{E}}{k_BT}\right)
    \end{equation}
    is the Boltzmann factor. A repulsive potential ($\mathcal{E}>0$) thus leads to a depletion of solute inside the layer ($K<1$), whereas an attractive potential ($\mathcal{E}<0$) leads to an accumulation ($K>1$). The corresponding equilibrium-interfacial excess density, given by Eq. (\ref{eq:-156}), is $n_1^{\phi|\text{eq}}=(K-1)n_1^{\text{b}|\text{eq}}$ inside the layer and zero elsewhere, and the amount of excess solute per unit area is characterized by the adsorption length 
    \begin{equation}
        l_\text{ads}=(K-1)\lambda,
    \end{equation}
    defined as the thickness of a film of bulk solution containing the same amount of solute per unit area as the interfacial excess.
    
    \textit{Thermodynamic forces (Fig. \ref{fig:-2}c).} Diffusiophoresis is caused by a solute diffusion flux across $\mathcal V$, which takes the diagonal form $\bm j_1=-(n_1^\text{eq}/\xi_1)\nabla\mathring\mu_1$, where $\xi_1$ is the friction coefficient of a solute particle. 	Since the solute is point-like ($\bar v_1=0$), $\mathring\mu_1$ is independent of the hydrodynamic pressure, and the solvent diffusion flux $\bm j_0=-\bar v_1\bm j_1/\bar v_0$ vanishes. The solvent thus occupies the entire volume of the solution, such that its modified chemical potential both balances the conservative part of the hydrostatic force density, as discussed in Section \ref{sec:-4}, and depends on the hydrodynamic pressure that enforces incompressible flow. Since colloidal diffusion is slow compared to solute diffusion, the solute continuity equation reduces to the diffusive steady-state condition
    \begin{equation}
    	\nabla\cdot\bm j_1=0.\label{eq:-82}
    \end{equation}
    The normal solute diffusion flux vanishes at the impermeable colloidal surface $r=R$, whereas $\mathring\mu_1$ and the normal solute diffusion flux are continuous at the outer boundary of the layer, $r=R+\lambda$. Far from the colloid, $\nabla\mathring\mu_1$ tends to $\bm G=k_BT\nabla n_1^\text{b}/n_1^{\text{b}|\text{eq}}$. Since $n_1^\text{eq}$ is uniform in each region, Eq. (\ref{eq:-82}) reduces to Laplace's equation there, and $\mathring\mu_1$ takes the dipolar form $\mathring\mu_1=\text{const}+f(r)\,\hat{\bm r}\cdot\bm G$, where $f(r)$ is a linear combination of $r$ and $r^{-2}$ in each region. Imposing the boundary conditions, we obtain to first order
    \begin{equation}
    	\nabla\mathring\mu_1=
    	\begin{cases}
    		\displaystyle A\bm\Phi\cdot\bm G, & R<r\le R+\lambda, \\[2ex]
    		\displaystyle \left[\bm I+B\frac{(R+\lambda)^3}{2r^3}\left(\bm I-3\hat{\bm r}\hat{\bm r}\right)\right]\cdot\bm G, & r>R+\lambda,
    	\end{cases}\label{eq:gradmu-step}
    \end{equation}
    with
    \begin{eqnarray}
    	&&\bm\Phi=\left(1+\frac{R^3}{2r^3}\right)\left(\bm I-\hat{\bm r}\hat{\bm r}\right)+\left(1-\frac{R^3}{r^3}\right)\hat{\bm r}\hat{\bm r},\nonumber\\
    	&&A=\frac{3(R+\lambda)^3}{D},\quad B=\frac{2(1-K)(R+\lambda)^3+(1+2K)R^3}{D},\nonumber\\
    	&&D=(2+K)(R+\lambda)^3+(1-K)R^3,\nonumber
    \end{eqnarray}
    where the tensor $\bm\Phi$ accounts for the vanishing normal flux at the colloidal surface. For $K=1$, the interfacial layer is absent, and Eq. (\ref{eq:gradmu-step}) reduces to the well-known dipolar solution around an impermeable sphere of radius $R$.\cite{brady2011particle} In the limit of an infinite potential ($K\to 0$), the solute is fully excluded from the layer, and $\mathring\mu_1$ is no longer constrained by Eq. (\ref{eq:-82}) inside it, but remains well defined as the limit of Eq. (\ref{eq:gradmu-step}), in accordance with its continuity within the solution. We then have $B=1$, such that the outer solution coincides with the dipolar solution around an impermeable sphere of radius $R+\lambda$.
    
    \textit{Auxiliary Stokes flow (Fig. \ref{fig:-2}d).} For an isolated spherical colloid, the friction coefficient and the Stokes flow tensor are given by Eqs. (\ref{eq:-73}) and (\ref{eq:-5}), respectively, for an arbitrary hydrodynamic slip parameter $a$.
    
    \textit{Diffusiophoretic velocity.} As shown in Section \ref{sec:-1}, the osmotic force density inside the solution can be evaluated using the reduced form of the hydrostatic force density, such that $\bm{\mathcal F}_\text{osm}=-\delta\left(n_1^{\text{eq}}\nabla\mathring\mu_1\right)=-n_1^{\phi|\text{eq}}\nabla\mathring\mu_1$, and hence
    \begin{equation}
    	\bm{\mathcal F}_\text{osm}=
    	\begin{cases}
    		\left(1-K\right)n_1^{\text{b}|\text{eq}}\nabla\mathring\mu_1, & R<r\le R+\lambda, \\[1ex]
    		0, & r>R+\lambda.
    	\end{cases}\label{eq:-122}
    \end{equation}
    We seek an explicit expression for the diffusiophoretic velocity $\bm v_\text{ph}$, written as
    \begin{equation}
    	\bm v_\text{ph}=\frac{R^2}{\eta}Mk_BT\nabla n_1^\text{b},\label{eq:-115}
    \end{equation}
    where $M$ is the rescaled diffusiophoretic mobility of the colloid. This velocity is given by the first term in Eq. (\ref{eq:-100}),
    \begin{equation}
    	\bm v_\text{ph}=\frac{1}{\xi}\int_{\tilde\nu}\left(\bm U^{*\dagger}-\bm I\right)\cdot\bm{\mathcal F}_\text{osm}\,d^3\bm r.\label{eq:-109}
    \end{equation}
    Using Eqs. (\ref{eq:-73}), (\ref{eq:-5}), (\ref{eq:gradmu-step}) and (\ref{eq:-122}) in Eq. (\ref{eq:-109}), and introducing the dimensionless layer thickness $x=\lambda/R$, we obtain
    \begin{equation}
    	M=\left(K-1\right)\Lambda\,\frac{x}{6a}\left[2x^2+3\left(2-a\right)x+6\left(1-a\right)\right],\label{eq:M-step}
    \end{equation}
    where the factor
    \begin{equation}
    	\Lambda=\frac{2(1+x)^3+1}{(2+K)(1+x)^3+1-K}
    \end{equation}
    accounts for the modification of the chemical potential gradient by the solute flux conducted through the interfacial layer, which tends to equalize the solute concentration around the colloid. 

    \subsection{Interpretation and limiting cases}\label{sec:IVC}
    
    The mobility given by Eq. (\ref{eq:M-step}) vanishes in the absence of an interfacial excess ($K=1$) and changes sign between repulsive and attractive potentials ($M<0$ for $K<1$ and $M>0$ for $K>1$), such that the colloid moves towards lower solute concentration for $K<1$ and towards higher solute concentration for $K>1$, over the entire range of hydrodynamic slip $2/3\le a\le 1$. In the excluded-volume limit ($K\to 0$), we have $\Lambda=1$, and Eq. (\ref{eq:M-step}) reduces to
    \begin{equation}
    	M=-\frac{x}{6a}\left[2x^2+3\left(2-a\right)x+6\left(1-a\right)\right].\label{eq:-125}
    \end{equation}
    
    \begin{table*}[t]
    	\centering
    	\renewcommand{\arraystretch}{2.0}
    	\begin{tabular}{lcl}
    		\hline\hline
    		\textbf{Limit} & \textbf{Reduced velocity $R^2M$} & \textbf{Agrees with} \\
    		\hline
    		$K=0$, $a=1$ &
    		$\displaystyle -\frac{\lambda^2}{2}-\frac{\lambda^3}{3R}$ &
    		Brady\cite{brady2021phoretic} \\
    		$x\ll 1$, $a=1$, leading order &
    		$\displaystyle \frac{l_\text{ads}\lambda}{2}$ &
    		Anderson\cite{anderson1989colloid} \\
    		$x\ll 1$, $a=1$, first order &
    		$\displaystyle \frac{l_\text{ads}\lambda}{2}\left(1-\frac{l_\text{ads}}{R}-\frac{\lambda}{3R}\right)$ &
    		Anderson \textit{et al.}\cite{anderson1982motion} \\
    		$x\to 0$ at fixed $l_\text{ads}/R$, $a=1$ &
    		$\displaystyle \frac{l_\text{ads}\lambda}{2}\left(1+\frac{l_\text{ads}}{R}\right)^{-1}$ &
    		Anderson and Prieve\cite{anderson1991diffusiophoresis} \\
    		$x,\ l_\text{slip}/R\ll 1$ &
    		$\displaystyle l_\text{ads}\left(\frac{\lambda}{2}+l_\text{slip}\right)\left(1+\frac{l_\text{ads}}{R}\right)^{-1}$ &
    		Ajdari and Bocquet\cite{ajdari2006giant} \\
    		$a=1$ &
    		$R^2\times$Eq. (\ref{eq:M-step}) with $a=1$ &
    		Marbach \textit{et al.}\cite{marbach2020local} \\
    		\hline\hline
    	\end{tabular}
    	\caption{Limiting cases of Eq. (\ref{eq:M-step}) and the corresponding results in the literature, expressed in terms of the reduced velocity $R^2M$, such that $\bm v_\text{ph}=R^2Mk_BT\nabla n_1^\text{b}/\eta$. The rows correspond, in order, to hard-sphere exclusion at arbitrary contact distance ($K=0$, no slip); the classical BLA result for thin layers at fixed $K$, with its first-order curvature correction; strongly adsorbing solutes in the absence of solute convection; thin layers with hydrodynamic slip at arbitrary $l_\text{slip}/\lambda$ and $l_\text{ads}/R$, corresponding to the velocity of a spherical particle derived by Ajdari and Bocquet in the absence of solute convection; and arbitrary $K$ and $x$ for a no-slip boundary condition. In the notation of these references, the first moment of the excess solute distribution is $L^*=\lambda/2$ for a step potential, and the dimensionless moment $H$ entering the curvature correction of Anderson \textit{et al.}\cite{anderson1982motion} takes the value $H=1/3$, giving the term $H\lambda/R$. The agreement with Marbach \textit{et al.}\cite{marbach2020local} is obtained by evaluating their general expression for a step potential.}
    	\label{tab:limits}
    \end{table*}
    
    Expanding Eq. (\ref{eq:M-step}) in the appropriate limits recovers the established results summarized in Table \ref{tab:limits}. For a no-slip boundary condition, Eq. (\ref{eq:-125}) coincides with the result obtained by Brady for hard-sphere exclusion at arbitrary contact distance,\cite{brady2021phoretic} and for thin layers, Eq. (\ref{eq:M-step}) recovers the classical BLA result,\cite{anderson1989colloid} which reduces to $\bm v_\text{ph}=-\lambda^2k_BT\nabla n_1^\text{b}/(2\eta)$ for steric exclusion ($K=0$), as well as the curvature correction derived by Anderson \textit{et al.}\cite{anderson1982motion} For strongly adsorbing solutes, the correction proportional to $l_\text{ads}/R$ need not be small, and the factor $\Lambda$ tends to $\left(1+l_\text{ads}/R\right)^{-1}$, recovering the result of Anderson and Prieve.\cite{anderson1991diffusiophoresis} The factor $\Lambda$ thus generalizes the effect of solute conduction through the interfacial layer to layers of arbitrary thickness. For thin layers with hydrodynamic slip, Eq. (\ref{eq:M-step}) recovers the diffusiophoretic velocity of a spherical particle derived by Ajdari and Bocquet in the absence of solute convection, which combines the enhancement factor $1+l_\text{slip}/L^*$ with the reduction by strong adsorption.\cite{ajdari2006giant} Finally, beyond the BLA, the general no-slip expression of Marbach \textit{et al.} for an arbitrary interaction range,\cite{marbach2020local} evaluated for a step potential, coincides with Eq. (\ref{eq:M-step}) for $a=1$. Eq. (\ref{eq:M-step}) thus unifies these predictions in a single closed-form expression and generalizes them to arbitrary layer thickness, potential strength and hydrodynamic surface slippage, thereby validating the continuum-statistical framework developed in Section \ref{sec:-7}. 

    \begin{figure*}[t]
	\centering
	\includegraphics[width=0.95\textwidth]{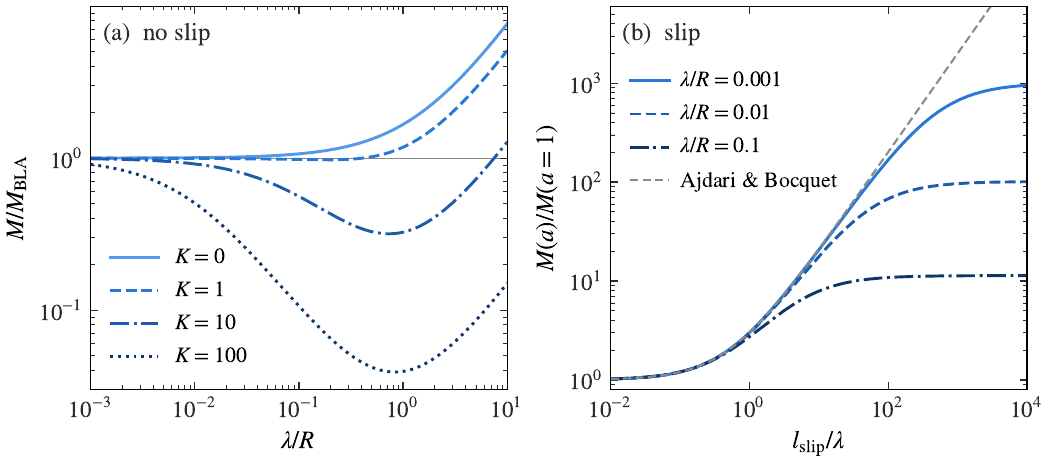}
    \caption{Trends of the diffusiophoretic mobility given by Eq. (\ref{eq:M-step}). (a) Ratio of the mobility to its BLA value for a no-slip boundary condition, $M/M_\text{BLA}=\Lambda\left(1+2\lambda/(3R)\right)$, as a function of the layer thickness for different Boltzmann factors $K$. (b) Enhancement of the mobility by hydrodynamic slip, $M(a)/M(a=1)$, which is independent of $K$, as a function of $l_\text{slip}/\lambda$ for different layer thicknesses. The dashed line shows the enhancement factor $1+l_\text{slip}/L^*$ obtained by Ajdari and Bocquet for a spherical particle, which implicitly assumes $l_\text{slip}\ll R$.\cite{ajdari2006giant}}
    	\label{fig:trends}
    \end{figure*}

    The trends of Eq. (\ref{eq:M-step}) beyond these limits are shown in Fig. \ref{fig:trends}. For a no-slip boundary condition, the ratio $M/M_\text{BLA}=\Lambda\left(1+2\lambda/(3R)\right)$ is independent of the sign of the interfacial excess. For excluded volume, the mobility exceeds its BLA value as the layer thickens, whereas for strongly adsorbing solutes, the factor $\Lambda$ reduces it by more than an order of magnitude when $\lambda$ becomes comparable to $R$ (Fig. \ref{fig:trends}a). The enhancement of diffusiophoresis by hydrodynamic slip follows the result of Ajdari and Bocquet as long as $l_\text{slip}\ll R$, but saturates when the slip length becomes comparable to the colloid radius, reaching approximately $R/\lambda$ for a perfect-slip boundary condition (Fig. \ref{fig:trends}b). Indeed, for thin layers at arbitrary $l_\text{slip}/R$, Eq. (\ref{eq:M-step}) yields
    \begin{equation}
    	R^2M=l_\text{ads}\left(1+\frac{l_\text{ads}}{R}\right)^{-1}\left[\frac{l_\text{slip}}{1+2l_\text{slip}/R}+\frac{\lambda}{2}\,\frac{1+4l_\text{slip}/R}{1+2l_\text{slip}/R}\right],\label{eq:vph-slip-sat}
    \end{equation}
    showing that the effective contribution of hydrodynamic slip is bounded by the finite size of the colloid. This geometric saturation is distinct from the saturation predicted by Ajdari and Bocquet for strongly adsorbing solutes, which results from solute convection at finite P\'eclet number,\cite{ajdari2006giant} and which is absent here, since solute convection is neglected.

	\section{Conclusion}

	In general, resolving the local continuum dynamics in colloidal suspensions is theoretically and computationally demanding as it requires solving multiple coupled continuity equations under moving boundary conditions. As a result, existing approaches often rely on linear-response models that treat diffusive transport, including force-free phoretic motion, in a largely phenomenological manner.

	Here, it has been shown that this phenomenology can be reduced under kinematic reversibility by considering the hydrodynamics of colloidal sedimentation. Within the presented continuum-statistical formulation, the Onsager reciprocal relations of colloidal transport arise naturally from the symmetry of the hydrodynamic response encoded in the Lorentz reciprocal theorem. Moreover, the resulting framework consistently accounts for the solvent's response to momentum relaxation, which balances the conservative part of the hydrostatic force density inside the solution, such that only equilibrium-interfacial excess densities contribute to it there, and fully resolves the coupling between thermodynamic forces and hydrodynamic flows beyond the boundary-layer approximation. As summarized in Fig. \ref{fig:-2} for a single colloid, the framework essentially consists of three independent and computationally tractable parts that represent the essential physical ingredients of a reciprocal treatment of colloidal diffusion: an equilibrium-statistical problem to determine the equation of state and local component distributions of the fluid (Fig. \ref{fig:-2}b), a steady-state problem to determine the local thermodynamic forces (Fig. \ref{fig:-2}c), and a hydrodynamic problem to compute the auxiliary Stokes flow under stationary boundary conditions (Fig. \ref{fig:-2}d).

	The framework has been validated by recovering the swimming velocity due to ciliary motion, and by applying it to diffusiophoresis caused by an interfacial step potential between the colloid and an ideal solute. The resulting closed-form expression for the diffusiophoretic mobility recovers established results as limiting cases, including the classical BLA result and its curvature correction, the effects of strong adsorption and hydrodynamic slip, and the results for hard-sphere exclusion and arbitrary interaction range, and generalizes them to arbitrary layer thickness, potential strength and hydrodynamic surface slippage. In particular, it shows that the enhancement of diffusiophoresis by hydrodynamic slip saturates once the slip length becomes comparable to the colloid radius.

	The underlying linear-response assumption is not unduly restrictive, as fluxes are almost universally taken to be linear in the non-equilibrium forces in advection–diffusion problems. A natural and promising direction for future work is the extension of this approach to rotational colloidal motion, including the coupling between the translational and rotational motion of nonspherical particles,\cite{brenner1967coupling,wegener1981diffusion,yuan2021electric} which could be described within the present framework by considering an auxiliary problem in which both a uniform force and a uniform torque are applied to the colloids. Furthermore, the approach could be applied numerically to non-ideal osmotic equations of state, including solutes of finite size and colloidal concentrations beyond the dilute limit, where correlations between colloid-specific forces, such as phoretic forces modified by neighboring microparticles, and the deviations of the auxiliary colloid motion from the mean sedimentation velocity may need to be included in the fluid-mechanical formulation (Section \ref{sec:-1}). At high colloidal volume fractions, however, it must be noted that static correlations may render the effective osmotic equation of state highly complex, and that dynamic arrest may eventually cause local structural relaxation times to diverge. Under these conditions, the underlying assumption of RTE is expected to fail, and a continuum-mechanical decoupling into hydrostatic and hydrodynamic stresses may no longer be justified.

    \section{Acknowledgments}
	The author acknowledges helpful discussions with Daan Frenkel.

    \section{Data availability statement}
    Data sharing is not applicable to this article as no new data were created or analyzed in this study.

    \appendix
    \section{Expressing the hydrostatic pressure gradient in terms of thermodynamic forces using the Gibbs-Duhem equation}\label{app:-1}

    Based on the Gibbs-Duhem equation, the pressure gradient can be written as
    \begin{equation}
        \nabla P=s\nabla T+\sum_i n_i\nabla\mu_i,
    \end{equation}
    where $s$ is the entropy density. Substituting Eq. (\ref{eq:-71}) for the chemical potential $\mu_i$, one finds
    \begin{equation}
        \nabla P=s\nabla T+\sum_i n_i\nabla\mathring \mu_i-\sum_i n_i\nabla\phi_i.\label{eq:-131}
    \end{equation}
    Decomposing the modified chemical potential gradient into $\nabla\mathring \mu_i=\left(\partial\mathring\mu_i/\partial T\right)_{\left\{n_j\right\}}\nabla T+\nabla_{T}\mathring\mu_i$, Eq. (\ref{eq:-131}) becomes  
    \begin{eqnarray}
        \nabla P&=&\left[Ts+T\sum_in_i\left(\frac{\partial\mathring\mu_i}{\partial T}\right)_{\left\{n_j\right\}}\right]\frac{\nabla T}{T}\nonumber\\
        &&+\sum_i n_i\nabla_T\mathring\mu_i-\sum_i n_i\nabla\phi_i.\label{eq:-143}
    \end{eqnarray}
    Since $\mathring\mu_i=\mu_i+\phi_i$, the Euler relation of the local thermodynamic state can be written as
    \begin{equation}
        Ts=h-\sum_i n_i\mathring\mu_i,\label{eq:-133}
    \end{equation}
    where $h$ is the net enthalpy density, which includes the interfacial potential energy density $\sum_i n_i\phi_i$. Using Eq. (\ref{eq:-133}) in Eq. (\ref{eq:-143}), one obtains
    \begin{equation}
        \nabla P=\left(h-\sum_in_i\bar H_i\right)\frac{\nabla T}{T}+\sum_i n_i\nabla_{T}\mathring\mu_i-\sum_i n_i\nabla\phi_i,\label{eq:-145}
    \end{equation}
    where
    \begin{equation}
        \bar H_i=-T^2\left(\frac{\partial}{\partial T}\frac{\mathring\mu_i}{T}\right)_{\left\{n_j\right\}}.
    \end{equation}
    The modified enthalpy density $\check h$ is then defined by $\check h=h-\sum_i n_i \bar H_i$. At constant number densities, the partial enthalpy $\bar H_i$ per particle of component $i$ coincides with the partial internal energy per particle.\cite{matubayasi2000super} Eq. (\ref{eq:-65}) is then recovered by evaluating Eq. (\ref{eq:-145}) to first order at constant hydrodynamic pressure, which removes $\nabla P_\text{d}$ from the left-hand side and its contributions $\bar v_i\nabla P_\text{d}$ from the chemical potential gradients on the right-hand side, where $\bar v_i$ is the partial volume per particle of component $i$. Since $\sum_i n_i\bar v_i=1$, these contributions cancel exactly, such that the result holds irrespective of whether the hydrodynamic pressure is uniform.

    \section{Consequences for earlier reciprocal treatments}\label{app:validity}

    Earlier reciprocal treatments of colloidal motion based on postulated Onsager reciprocity led to a fluid-mechanical form of the colloidal diffusion flux [Eq. (62) of Ref.~\onlinecite{burelbach2019particle}] that agrees with Eq. (\ref{eq:-100}), including the contributions of the colloidal osmotic pressure gradient and of buoyant sedimentation, apart from a spurious contribution to the phoretic term. This phoretic term was subsequently used in Ref.~\onlinecite{burelbach2019linear} and generalized to the linear and angular velocities of a single self-diffusiophoretic Janus particle, where it again contains a spurious contribution. Both spurious contributions stem from the role attributed to the solvent. 
    
    For a planar surface subjected to a uniform thermodynamic force along it, the solution far from the surface is homogeneous and at hydrostatic equilibrium, which fixes the gradient of the modified chemical potential of the solvent, and the hydrostatic force density inside the interfacial layer reduces to the tangential gradient of the local pressure relative to its bulk value.\cite{ganti2017molecular} Since the interfacial potentials only vary normal to the surface, this tangential gradient coincides with $\nabla_{\left\{\phi_j\right\}}P_\text{s}$ given by Eq. (\ref{eq:-64}), and only equilibrium-interfacial excess densities contribute to it. In the aforementioned reciprocal treatments of colloidal motion, where the thermodynamic forces are in general non-uniform, the solvent was likewise assumed to balance the conservative part of the hydrostatic force density only in the bulk of the system,\cite{burelbach2019particle} or only its spherically symmetric part.\cite{burelbach2019linear} As established in Section \ref{sec:-4}, however, the solvent balances this conservative part locally everywhere inside the solution, such that only equilibrium-interfacial excess densities contribute to the hydrostatic force density there, as expressed by Eq. (\ref{eq:-67}). When the thermodynamic forces are perturbed by the colloid or generated by its surface activity, they vary on the colloidal scale, and the earlier prescriptions retain contributions of the bulk densities coupled to the local variation of the thermodynamic forces, which vanish for uniform thermodynamic forces, but are otherwise spurious.

    In Ref.~\onlinecite{burelbach2019particle}, these contributions stem from the perturbation of the reservoir force density given by Eq. (57) of that work, and give rise to the second term in Eq. (81) of that work. Consequently, no diffusiophoretic motion should occur in the absence of an interfacial excess, in contrast to Eqs. (86)-(88) of Ref.~\onlinecite{burelbach2019particle}, and the contributions of the bulk densities to the electrophoretic and thermophoretic mobilities, given by the last term in Eq. (95) and by the term proportional to the colloidal volume fraction in Eq. (103) of that work, are absent. In Ref.~\onlinecite{burelbach2019linear}, only the interfacial excess densities should contribute to the active linear and angular velocities of a Janus particle given by Eqs. (22) and (23) of that work, such that no active motion occurs in the absence of an interfacial solute excess, in contrast to the contribution $v_0$ given by Eq. (52) of that work.

    All interfacially driven predictions of these treatments remain valid. In particular, the first term in Eq. (81) of Ref.~\onlinecite{burelbach2019particle} gives the phoretic force on a spherical colloid for arbitrary hydrodynamic slip, interfacial layer thickness and non-uniform thermodynamic forces, and thus generalizes the Henry function for electrophoresis and the Ruckenstein term for thermophoresis. Since the contributions of the bulk densities to the electrophoretic mobility vanish by electroneutrality whenever all solutes share the same diffusive permeability ratio across the hydrodynamic boundary, the recovery of the Henry function [Eqs. (98) and (99) of that work] is unaffected, as is the recovery of the Ruckenstein term [Eqs. (105) and (106) of that work]. Likewise, the active linear velocity of half-capped Janus particles due to an interfacial solute excess [Eqs. (53) and (55) of Ref.~\onlinecite{burelbach2019linear}], their active angular velocity [Eq. (71) of that work] and their far-field diffusiophoretic interactions [Eqs. (60)-(66) of that work] remain valid, where the angular velocities rely on the reciprocal relations for rotational motion postulated in Ref.~\onlinecite{burelbach2019linear}.

\bibliographystyle{unsrt}
\bibliography{Ref}

\end{document}